\documentclass[11pt]{article}

\usepackage{graphicx}
\usepackage{amsfonts}
\usepackage{SIGACTNews}
\usepackage{indentfirst}
\usepackage{amsmath}
\usepackage{amsthm}
\usepackage{amssymb}
\usepackage[numbers]{natbib}
\usepackage{hyperref}
\usepackage{cleveref}
\usepackage{indentfirst}
\usepackage{color}
\usepackage{multicol}
\usepackage[boxed,linesnumbered]{algorithm2e}
\usepackage{bm}

\usepackage{times}

\definecolor{corlinks}{RGB}{0,0,150}
\definecolor{cormenu}{RGB}{0,0,150}
\definecolor{corurl}{RGB}{0,0,150}

\hypersetup{
	colorlinks=true,
	urlcolor=corlinks,
	linkcolor=corlinks,
	menucolor=cormenu,
	citecolor=corlinks,
	pdfborder= 0 0 0
}

\definecolor{blue-violet}{rgb}{0.54, 0.17, 0.89}

\title{SIGACT News Complexity Theory Column 118}
\author{Ben Lee Volk\\
Efi Arazi School of Computer Science\\Reichman University\\
Herzliya, Israel}
\date{
\includegraphics[height=37mm]{benlee.jpg}
}
\newcommand{\eqdef}{\triangleq}

\newcommand{\poly}{\mathsf{poly}}

\renewcommand{\P}{\mathsf{P}}
\newcommand{\NP}{\mathsf{NP}}
\newcommand{\FP}{\mathsf{FP}}
\newcommand{\coNP}{\mathsf{coNP}}

\newcommand{\PV}{\mathsf{PV}_1}
\newcommand{\LPV}{\mathcal{L}_{\mathsf{PV}}}
\newcommand{\LBuss}{\mathcal{L}_\mathsf{B}}
\newcommand{\dWPHP}{\mathsf{dWPHP}}
\newcommand{\SOT}{\mathsf{S}^1_2}
\newcommand{\TZT}{\mathsf{T}^0_2}
\newcommand{\TOT}{\mathsf{T}^1_2}
\newcommand{\TTT}{\mathsf{T}^2_2}
\newcommand{\APC}{\mathsf{APC}}

\newcommand{\EF}{e\mathcal{F}}
\newcommand{\Prf}{\mathsf{Proof}}
\newcommand{\Sat}{\mathsf{Sat}}

\newcommand\pnp{\varphi_{\mathsf{P} = \mathsf{NP}}}
\newcommand\LB{\mathsf{LB}_k^{\textit{a.e.}}}
\newcommand\upci{{\mathsf{UB}^{i.o.}_{k}}}

\newtheorem{theorem}{Theorem}[section]
\newtheorem{lemma}[theorem]{Lemma}
\newtheorem{fact}[theorem]{Fact}
\newtheorem{claim}[theorem]{Claim}

\newtheorem{problem}[theorem]{Open Problem}

\theoremstyle{definition}
\newtheorem{definition}[theorem]{Definition}

\theoremstyle{remark}

\begin{document}

\clearpage

\newpage

\begin{center}
{\Large  
SIGACT News Complexity Theory Column\vspace{0.3cm}\\{ \bf Meta-Mathematics of Computational Complexity Theory}\vspace{0.2cm}}\\

\medskip
{\Large 
Igor C. Oliveira}\footnote{Department of Computer Science, University of Warwick, UK. Email: \textbf{\texttt{igor.oliveira@warwick.ac.uk}}.}

\bigskip 

\vspace{-0.1cm}

\includegraphics[height=28mm]{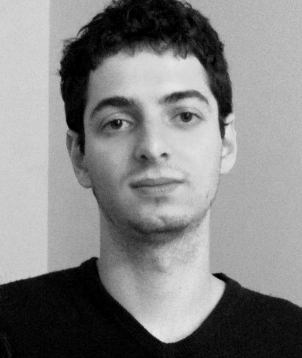}

\vspace{-0.1cm}

\end{center}

\begin{abstract}
We survey results on the formalization and independence of mathematical statements related to major open problems in computational complexity theory. Our primary focus is on recent findings concerning the (un)provability of complexity bounds within theories of bounded arithmetic. This includes the techniques employed and related open problems, such as the (non)existence of a feasible proof that $\mathsf{P} = \mathsf{NP}$.  
\end{abstract}

\vspace{-0.6cm}

\small

\setcounter{tocdepth}{3}

\tableofcontents

\normalsize

\section{Introduction}\label{sec:introduction}

The investigation of the inherent complexity of computational tasks is a central research direction in theoretical computer science. While unconditional results are known in a variety of restricted contexts (i.e.,~with respect to weak models of computation), despite significant efforts, several central questions of the field remain wide open. Prominent examples include the relation between complexity classes $\mathsf{P}$ and $\mathsf{NP}$,  understanding the power of non-uniform Boolean circuits, and bounding the length of proofs in  propositional proof systems such as Frege and  extended Frege.

The investigation of the difficulty of settling these problems has long been an important and influential area of research by itself (e.g.,~barrier results such as \citep{DBLP:journals/siamcomp/BakerGS75, DBLP:journals/jcss/RazborovR97, DBLP:journals/toct/AaronsonW09, DBLP:journals/jacm/ChenHOPRS22}). Unfortunately, these results tend to be ad-hoc and do not consider a standard and robust notion of proof. In order to build a general theory, several works have considered provability in the usual sense of mathematical logic. Most importantly, this enables a deeper investigation of complexity theory that considers not only the running time of a program or the size of a circuit but also the feasibility of proving their existence and correctness. In particular, we can explore the fundamental question of what can and cannot be feasibly computed, along with the meta-question of what lower and upper bounds can and cannot be feasibly proven.

A fundamental goal of this research is to\\

\vspace{-0.2cm}

\noindent $(\star)$ identify a suitable logical theory capable of formalizing most, if not all, known results in algorithms and complexity, and determine whether the major open problems mentioned above are provable or unprovable within this theory.\footnote{As we elaborate in \Cref{sec:unprov}, the unprovability of a statement is equivalent to the consistency of its negation, which can be at least as important.}\\

\vspace{-0.2cm}

Although we are still far from reaching this goal, progress has been made in understanding the (un)provability of statements concerning the complexity of computations within certain fragments of Peano Arithmetic, collectively known as Bounded Arithmetic. These theories are designed to capture proofs that manipulate and reason with concepts from a specified complexity class. For instance, a proof by induction whose inductive hypothesis can be expressed as an $\NP$ predicate is one such example. The earliest theory of this kind was $\mathsf{I}\Delta_0$, introduced by Parikh \citep{Parikh}, who explored the intuitive concept of feasibility in arithmetic and addressed the infeasibility of exponentiation. The relationship between Parikh's theory and computational complexity was fully recognized and advanced by Paris and Wilkie in a series of influential papers during the 1980s (see \citep{DBLP:journals/apal/WilkieP87}). Other significant theories include Cook’s theory $\PV$ \citep{Coo75}, which formalizes polynomial-time reasoning; Jeřábek’s theory $\APC_1$ \citep{Jerabek04, Jerabek-phd, Jerabek07}, which extends $\PV$ by incorporating the dual weak pigeonhole principle for polynomial-time functions and formalizes probabilistic polynomial-time reasoning; and Buss’s theories $\mathsf{S}^i_2$ and $\mathsf{T}^i_2$ \citep{Buss}, which include induction principles corresponding to various levels of the polynomial-time hierarchy.

These theories are capable of formalizing  advanced results. For instance, it is known that $\PV$ can prove the PCP Theorem \citep{DBLP:journals/corr/Pich14}, while $\APC_1$ can establish several significant circuit lower bounds \citep{DBLP:journals/apal/MullerP20}, including monotone circuit lower bounds for $k$-Clique and bounded-depth circuit lower bounds for the Parity function. Further examples include the explicit construction of expander graphs \citep{DBLP:journals/apal/BussKKK20} and the correctness of randomized polynomial-time matching algorithms \citep{DBLP:journals/corr/abs-1103-5215}, among many others.

Given the expressive power of these theories, even if we are not yet able to establish a breakthrough result of the magnitude of $(\star)$, determining the (un)provability of complexity bounds of interest in theories of bounded arithmetic still represents significant progress towards our understanding of the power and limits of feasible computations and proofs. This survey aims to provide an introduction to some of these results, the underlying techniques, and related open problems. While our primary focus is on recent developments, in order to provide a broader perspective we also cover some classical results. Due to space limitations, the survey is not exhaustive, and several references had to be omitted (although some recent developments are mentioned in \Cref{sec:recent_developments}).

\section{Preliminaries}

\subsection{Complexity Theory}\label{sec:prelim_complexity}

We will rely on a few additional standard definitions from complexity theory, such as basic complexity classes, Boolean circuits and formulas, and propositional proof systems. These can be found in textbooks such as 
\citep{AB09} and \citep{krajicek_2019}. Below we only establish notation and review a classical result that offers a convenient way to talk about polynomial-time computations in some logical theories.

We use $\mathsf{SIZE}[s]$ to denote the set of languages computed by Boolean circuits of size $s(n)$.

In theoretical computer science, one typically considers functions and predicates that operate over binary strings. This is equivalent to operations on integers, by identifying each non-negative integer with its binary representation. Let $\mathbb{N}$ denote the set of non-negative integers. For $a \in \mathbb{N}$, we let $|a| \eqdef \lceil \log_2(a+1) \rceil$ denote the length of the binary representation of $a$. For a constant $k \geq 1$, we say that a function $f \colon \mathbb{N}^k \to \mathbb{N}$ is computable in polynomial time if $f(x_1, \ldots, x_k)$ can be computed in time polynomial in $|x_1|, \ldots, |x_k|$. (For convenience, we might write $|\vec{x}| \eqdef |x_1|, \ldots, |x_k|$.)  Recall that $\FP$ denotes the set of polynomial time functions. While the definition of polynomial time refers to a machine model, $\FP$ can also be introduced in a machine independent way as the closure of a set of base functions under \emph{composition} and \emph{limited recursion on notation}. In more detail, we can consider the following class $\mathcal{F}$ of base functions:
$$
c(x) \eqdef 0, \quad s(x) \eqdef x + 1, \quad a(x) \eqdef \lfloor x/2 \rfloor, \quad d(x) \eqdef 2 \cdot x,  \quad \pi^i_\ell(x_1, \ldots, x_\ell) \eqdef x_i, \quad x \# y \eqdef 2^{|x| \cdot |y|},
$$
\vspace{-0.8cm}
\begin{align}
    x \leq y  \eqdef  
    \begin{cases}
      1  \quad \text{if}~x \leq y \\
      0   \quad       \text{otherwise,}
    \end{cases}
&  \quad \quad \quad
    \mathsf{Choice}(x,y,z)  \eqdef  \begin{cases}
      y \quad  \text{if}~x > 0 \\
      z  \quad        \text{otherwise.}
    \end{cases}
    \nonumber 
\end{align}
We say that a function $f(\vec{x},y)$ is defined from functions $g(\vec{x})$, $h(\vec{x},y,z)$, and $k(\vec{x},y)$ by \emph{limited recursion on notation} if
\begin{eqnarray}
f(\vec{x},0) & = & g(\vec{x}) \nonumber \\
f(\vec{x},y) & = & h(\vec{x},y,f(\vec{x},\lfloor y/2 \rfloor)) \nonumber \\
f(\vec{x},y) & \leq & k(\vec{x}, y) \nonumber
\end{eqnarray}
for every sequence $(\vec{x}, y)$ of natural numbers. 
Cobham \citep{Cob64} proved that $\FP$ is the least class of functions that contains $\mathcal{F}$ and is closed under composition and limited recursion on notation.

\subsection{Theories of Bounded Arithmetic}

Bounded arithmetic has a long and rich history (see \citep{buss-survey} for an introduction, and \citep{hajek2017metamathematics, Krajicek-book, cook_nguyen_2010} for a detailed treatment). The correspondence between the
theories and complexity classes manifests in multiple ways. For instance,  \emph{witnessing results} show that every provably total function in a given theory $\mathsf{T}_\mathcal{C}$ (i.e.,~when $\forall x~\exists! y~\psi(x,y)$ is provable, for certain formulas $\psi$) is computable within the corresponding complexity class $\mathcal{C}$ (i.e., the function $y = f(x)$ is in $\mathcal{C}$). There is also a close connection between theories of bounded arithmetic and propositional proof systems, e.g., \emph{propositional translations} between proofs of certain sentences in $\PV$ or $\SOT$ and polynomial-size proofs in the extended Frege proof system of the corresponding propositional formulas. We review some related results in \Cref{sec:prelim_witnessing} and \Cref{sec:prelim_prop_proofs}, respectively. In this section, we provide an overview of some widely investigated theories of bounded arithmetic: $\PV$, $\SOT$, $\TOT$, and $\APC_1$. We assume basic familiarity with first-order logic. Results claimed below without reference can be found in \citep{Krajicek-book}.

\subsubsection{$\PV$}\label{sec:PV}

$\PV$ \citep{Coo75} (see also \citep{KrajicekPT91})  is a first-order theory whose intended model is the set  $\mathbb{N}$ of natural numbers, together with the standard interpretation for constants and functions symbols such as $0, +, \times$, etc. The vocabulary (language) of $\PV$, denoted $\mathcal{L}_{\PV}$, contains a function symbol for each polynomial-time algorithm $f \colon \mathbb{N}^k \to \mathbb{N}$ (where $k$ is any constant). These function symbols, and the axioms defining them, are obtained through Cobham's characterization of polynomial-time functions discussed in \Cref{sec:prelim_complexity}. 

$\PV$ also postulates an induction axiom scheme that simulates binary search, and one can show that it admits induction over quantifier-free formulas (i.e.,~polynomial-time predicates). We discuss induction axioms in more detail in \Cref{sec:Busss_theories}.  

We will use later in the text that $\PV$ admits a formulation where all axioms are universal formulas (i.e.,~$\forall \vec{x}\, \phi(\vec{x})$, where $\phi$ is a quantifier-free formula). In other words, $\PV$ is a \emph{universal theory}.  

While the details of the definition of
$\PV$ are fairly technical (see, e.g., the longer overview in \citep{CLO24a} or the exposition in \citep{Krajicek-book}), such details are often not needed. In particular, $\PV$ has an equivalent  formalization that does not require Cobham's result \citep{jerabek:sharply-bounded}.

\subsubsection{$\SOT$, $\TOT$, and Beyond}\label{sec:Busss_theories}

While $\PV$ can be related  to polynomial-time computations and feasible proofs, Buss \citep{Buss} introduced a hierarchy of theories with close ties to the different levels of the polynomial hierarchy. To specify the theories, we will need a few definitions.

The language $\LBuss$ of these theories contains the predicate symbols $=$ and $\leq$, the constant symbols $0$ and $1$, and function symbols $S$ (successor), $+$, $\cdot$, $\lfloor x/2 \rfloor$, $|x|$ (interpreted as the length of $x$ as in \Cref{sec:prelim_complexity}), and $\#$ (``smash''; interpreted as $x \# y = 2^{|x| \cdot |y|}$).

A \emph{bounded quantifier} is a quantifier of the form $Qy \leq t$, where $Q \in \{\exists, \forall\}$ and $t$ is a term not involving $y$. Similarly, a \emph{sharply bounded quantifier} is one of the form $Qy \leq |t|$. Formally, such quantifiers are simply abbreviations. For instance,
\begin{eqnarray}
    \forall y \leq t(\vec{x})\;\varphi(\vec{x},y) & \eqdef & \forall y\,(y \leq t(\vec{x}) \rightarrow \varphi(\vec{x},y)),~\text{and} \nonumber \\
     \exists y \leq t(\vec{x})\;\varphi(\vec{x},y) & \eqdef & \exists y\,(y \leq t(\vec{x}) \wedge \varphi(\vec{x},y))\,. \nonumber
\end{eqnarray}
A formula where each quantifier appears bounded (resp.,~sharply bounded) is said to be a bounded (resp.,~sharply bounded) formula. It is not hard to show that every sharply bounded formula defines a polynomial-time predicate over the standard model $\mathbb{N}$ under its usual operations. On the other hand, bounded quantifiers allow us to define predicates in $\NP$, $\coNP$, and beyond. 

We can introduce a hierarchy of formulas by counting alternations of bounded quantifiers. The class $\Pi^b_0 = \Sigma^b_0$ contains the sharply bounded formulas. We then recursively define, for each $i \geq 1$, the classes $\Sigma^b_i$ and $\Pi^b_i$ according to the quantifier structure of the sentence, ignoring the appearance of sharply bounded quantifiers. For instance, if $\varphi \in \Sigma^b_0$ and $\psi \eqdef \exists y\leq t(\vec{x})\;\varphi(y,\vec{x})$, then $\psi \in \Sigma^b_1$ (see, e.g., \citep{Krajicek-book} for the technical details in the general case). As alluded to above, it is known that, for each $i \geq 1$, a predicate $P(\vec{x})$ is in $\Sigma^p_i$ (the $i$-th level of the polynomial hierarchy) if and only if  there is a $\Sigma^b_i$-formula that agrees with it over $\mathbb{N}$.

The theories introduced by Buss share a common set BASIC of finitely many axioms postulating the expected arithmetic behavior of the constants, predicates, and function symbols, e.g., $x + y = y + x$ and $|1| = 1$ (see, e.g.,~\citep[Page 68]{Krajicek-book} for the complete list). The only difference among the theories is the kind of induction axiom scheme that each of them postulates.

\paragraph{Theory $\mathsf{T}^1_2$.}  This is a theory in the language $\LBuss$ extending BASIC by the induction axiom $\mathsf{IND}$
$$
\varphi(0) \wedge \forall x\,(\varphi(x) \rightarrow \varphi(x+1)) \rightarrow\;\forall x\,\varphi(x)
$$
for all $\Sigma^b_1$-formulas $\varphi(a)$. The formula $\varphi(a)$ may contain other free variables in addition to $a$.\\

We say that $\TOT$ supports induction for $\NP$ predicates. Intuitively, this means that we can aim to prove a result in $\TOT$ by induction, provided the induction hypothesis is defined by a predicate computable in $\NP$.  This definition can be extended to a theory that postulates induction for $\Sigma^b_i$-formulas, which gives rise to the theory $\mathsf{T}^i_2$.

\paragraph{Theory $\mathsf{S}^1_2$.} This is a theory in the language $\LBuss$ extending BASIC by the polynomial induction axiom $\mathsf{PIND}$
$$
\varphi(0) \wedge \forall x\,(\varphi(\lfloor x/2 \rfloor) \rightarrow \varphi(x)) \rightarrow\;\forall x\,\varphi(x)
$$
for all $\Sigma^b_1$-formulas $\varphi(a)$. The formula $\varphi(a)$ may contain other free variables in addition to $a$.\\ 

Intuitively, polynomial induction reduces the proof of $\varphi(x)$ to proving $\varphi(\lfloor x/2 \rfloor)$. Unlike the standard induction axiom, this approach allows us to reach the base case in just $\mathsf{poly}(n)$ steps when starting with an integer $x$ represented by $\mathsf{poly}(n)$ bits. This has implications for the efficiency of translating certain proofs in $\mathsf{S}^1_2$ into sequences of propositional proofs and for the extraction of polynomial-time algorithms from proofs (see \Cref{sec:prelim_witnessing} and  \Cref{sec:prelim_prop_proofs}) . Analogously to $\mathsf{T}^i_2$, we can define the theories $\mathsf{S}^i_2$ via polynomial induction for $\Sigma^b_i$-formulas. 

It is known that $\PV$ is essentially equivalent to $\TZT$ under an appropriate vocabulary and axioms \citep{jerabek:sharply-bounded}, and that $\mathsf{S}^i_2 \subseteq \mathsf{T}^{i}_2 \subseteq \mathsf{S}^{i+1}_2$ for every $i \geq 1$.

When stating and proving results in $\SOT$, it is  convenient to employ a more expressive vocabulary under which any polynomial-time function can be easily described. Moreover, it is possible to achieve this in a \emph{conservative} way, i.e., without  increasing the power of the theory. In more detail, let $\Gamma$ be a set of $\LBuss$-formulas. We say that a polynomial-time function $f \colon \mathbb{N}^k \to \mathbb{N}$ is $\Gamma$\emph{-definable} in $\SOT$ if there is a formula $\psi(\vec{x},y) \in \Gamma$ for which the following conditions hold:
\begin{itemize}
    \item[(\emph{i})] For every $a \in \mathbb{N}^k$, $f(\vec{a}) = b$ if and only if $\mathbb{N} \models \varphi(\vec{a},b)$.
    \item[(\emph{ii})] $\SOT \vdash \forall \vec{x}\,(\exists y\,(\varphi(\vec{x},y) \wedge \forall z\,(\varphi(\vec{x},z) \rightarrow y = z))\,.$ 
\end{itemize}
Every function $f \in \FP$ is $\Sigma^b_1$-definable in $\SOT$. By adding all functions in $\FP$ to the vocabulary of $\SOT$ and by extending $\SOT$ with their defining axioms (i.e.,~$\forall x\,\varphi(x,f(x))$), we obtain a theory $\SOT(\LPV)$ that can refer to polynomial-time predicates using quantifier-free formulas. $\SOT(\LPV)$ proves the polynomial induction scheme for both $\Sigma^b_1$-formulas and $\Pi^b_1$-formulas in the extended vocabulary.  $\SOT(\LPV)$ is conservative over $\SOT$, in the sense that any $\LBuss$-sentence provable in $\SOT(\LPV)$ is also provable in $\SOT$. 

A $\forall \Sigma^b_i$-sentence is simply a sentence $\psi = \forall \vec{x}\,\varphi(\vec{x})$ where $\varphi \in \Sigma^b_i$. Every $\forall \Sigma^b_1$-sentence provable in $\SOT(\LPV)$ is also provable in $\PV$. In other words, $\SOT(\LPV)$ is $\forall \Sigma^b_1$-conservative over $\PV$. On the other hand, it is known that if $\SOT(\LPV) = \PV$, then the polynomial-time hierarchy collapses.

\subsubsection{$\APC_1$}

In order to formalize probabilistic methods and randomized algorithms, Je\v r\'abek \citep{Jerabek04, Jerabek-phd, Jerabek07} formulated the theory $\APC_1$ (this terminology is from \citep{DBLP:journals/jsyml/BussKT14}) by extending $\PV$ with the \emph{dual Weak Pigeonhole Principle} ($\dWPHP$) for $\PV$ functions:\footnote{The $\dWPHP$ axiom scheme is also referred to as the surjective Weak Pigeonhole Principle in some references.}
$$
\APC_1 \eqdef \PV \cup \{\dWPHP(f) \mid f \in \LPV\}.
$$
Informally, each sentence $\dWPHP(f)$ postulates that, for every length $n = |N|$, there is $y < (1 + 1/n)\cdot N$ such that $f(x) \neq y$ for every $x < N$. 

It is known that the dual Weak Pigeonhole Principle for polynomial-time predicates can be proved in $\TTT$ \citep{MACIEL2002843}, and consequently $\APC_1 \subseteq \TTT(\LPV)$.

\section{Auxiliary Definitions and Results}

\subsection{Witnessing Theorems}\label{sec:prelim_witnessing}

Suppose a sentence $\psi$ of a certain syntactic form  admits a proof in a theory $T$ over a vocabulary $\mathcal{L}$. A witnessing theorem allows us to extract computational information from any such proof, by showing that an existential quantifier in $\psi$ can be witnessed by $\mathcal{L}$-terms. The simplest example of such a result is stated next.

\begin{theorem}[Herbrand's Theorem~(see, e.g.,~\citep{DBLP:conf/lcc/Buss94, mckinley2010sequent})]\label{thm:Herbrand}
    Let $T$ be a universal theory over a vocabulary $\mathcal{L}$. Let $\varphi(x,y)$ be a quantifier-free $\mathcal{L}$-formula, and suppose that $T \vdash \forall x\,\exists y\,\varphi(x,y)\,.$ There  is a constant $k \geq 1$ and $\mathcal{L}$-terms $t_1(x), \ldots, t_k(x)$ such that
    $$
T \vdash \varphi(x,t_1(x)) \vee \varphi(x,t_2(x)) \vee \ldots \vee \varphi(x,t_k(x))\,.
    $$
\end{theorem}

As an immediate consequence, if we apply  \Cref{thm:Herbrand} to $T \eqdef \PV$, we obtain $\LPV$-terms (corresponding to polynomial-time functions over $\mathbb{N}$) such that, given $a \in \mathbb{N}$, at least one of them produces a witness $b \in \mathbb{N}$ such that $\mathbb{N} \models \varphi(a,b)$.

Next, we consider the provability of more complex sentences in a universal theory.

\begin{theorem}[KPT Theorem~\citep{KrajicekPT91}]\label{thm:KPT}
Let $T$ be a universal theory with vocabulary $\mathcal{L}$, $\varphi(w,u,v)$ be a quantifier-free $\mathcal{L}$-formula, and suppose that $T\, \vdash\, \forall w\,\exists u\,\forall v\, \varphi(w,u,v).$ Then there exist a constant $k \geq 1$ and $\mathcal{L}$-terms $t_1, \ldots, t_k$ such that
$$
T\, \vdash\, \varphi(w,t_1(w), v_1) \vee \varphi(w,t_2(w,v_1),v_2) \vee \ldots \vee \varphi(w, t_k(w, v_1, \ldots, v_{k-1}), v_k)\ ,
$$
where the notation $t_i(w, v_1, \ldots, v_{i-1})$ indicates that these are the only variables occurring in $t_i$.   
\end{theorem}

 \Cref{thm:KPT} has a natural interpretation as an interactive game with finitely many rounds, which we revisit in \Cref{sec:LEARN_unprov} in the context of the provability of circuit upper bounds.

A similar form of \Cref{thm:KPT} holds under the provability of a $\forall \exists \forall \exists$-sentence (see, e.g.,~\citep{CKKOT24} for a concrete application in the context of circuit lower bounds). In contrast, there is no straightforward analogue of the KPT Theorem for a larger number of quantifier alternations. In this case, more general formulations are needed, such as the ones considered in  \citep{pudlak2006consistency, DBLP:journals/jsyml/BussKT14, LO23}.

It is also possible to establish witnessing theorems for theories that are not universal. This can be done either by first transforming the theory into a universal theory through the inclusion of new function symbols and quantifier elimination, or via direct approaches (see, e.g.,~\citep[Section 7.3]{Krajicek-book}). Another  example is Buss's Theorem for $\SOT$, which can be used to show that every $\forall \Sigma^b_1$-sentence provable in  $\SOT(\LPV)$ is also provable in $\PV$. This has two implications. First, we can combine this result with \Cref{thm:Herbrand}, which yields polynomial-time algorithms from proofs of $\forall \Sigma^b_1$-sentences in $\SOT(\LPV)$. Second, this means that in some situations we can establish the provability of a sentence in $\PV$ using the more convenient theory $\SOT(\LPV)$ (see \Cref{sec:formula_lb_formalization} for an example).

\subsection{Bounded Arithmetic and Propositional Proofs}\label{sec:prelim_prop_proofs}

In this section, we explain a connection between $\PV$ and the extended Frege proof system discovered by \citep{Coo75}. In short, it says that if a universal $\LPV$-sentence $\phi(x)$ is provable in $\PV$, then there is a translation of $\phi(x)$ into a sequence $\{G_n\}_{n \geq 1}$ of propositional formulas $G_n(p_1, \ldots, p_n)$ such that each $G_n$ has an extended Frege proof $\pi_n$ of size polynomial in $n$.\footnote{Conceptually, this is analogous to the translation of a polynomial-time Turing machine $M$ into a sequence $\{C_n\}_{n \geq 1}$ of polynomial-size Boolean circuits, one for each input length $n$.} 

First, we review some concepts and fix notation, deferring the details to a standard textbook (e.g.,~\citep{krajicek_2019}). Recall that a propositional formula $G(p_1, \ldots, p_n)$ is formed using variables $p_1, \ldots, p_n$, constants $0$ and $1$, and logical connectives $\wedge$, $\vee$, and $\neg$. A \emph{Frege} ($\mathcal{F}$) proof system is a ``textbook'' style proof system for propositional logic. It can be formulated as a finite set of axiom schemes together with the modus ponens rule. $\mathcal{F}$ is known to be sound and complete. The \emph{size} of a Frege proof is the total number of symbols occurring in the proof. In the \emph{extended Frege} ($e\mathcal{F}$) proof system, we also allow repeated subformulas appearing in a proof to be abbreviated via new variables.

\paragraph{Cook's Translation \citep{Coo75}.} Let $\varphi$ be a universal $\LPV$-sentence of the form $\varphi \eqdef \forall x\,\psi(x)$, where $\psi(x)$ is a quantifier-free formula. Cook \citep{Coo75} established that if $\varphi$ is provable in $\PV$, then there is a sequence $\{G_n\}_{n \geq 1}$ of propositional tautologies such that
\begin{itemize}
    \item[--] Each $G_n(p_1, \ldots, p_n)$ is a polynomial-size formula.\footnote{We note that $G_n(p_1, \ldots, p_n)$ might contain auxiliary variables beyond $p_1, \ldots, p_n$.}
    \item[--] $G_n$ encodes that $\psi(x)$ is true whenever $|x| \leq n$, i.e., over all integers encoded as $n$-bit strings.
    \item[--] $G_n$ admits polynomial-size $e\mathcal{F}$-proofs.
    \item[--] Moreover, the existence of polynomial-size $e\mathcal{F}$-proofs for each $G_n$ is provable in $\PV$. (We will need this additional property of the translation in \Cref{sec:eF_lbs}.)
\end{itemize}
For a formula  $\psi(x)$ as above, we often write $||\psi||_n$ to denote the corresponding propositional formula over inputs of length $n$.\\

For more information about the relation between proofs in bounded arithmetic and propositional proofs, including additional examples of propositional translations,  we refer to \citep{beyersdorff2009correspondence, krajicek_2019}.

\subsection{Cuts of Models of Bounded Arithmetic}\label{sec:cuts}

Many fundamental results in bounded arithmetic are established using model-theoretic techniques (see, e.g., the exposition of Parikh's Theorem in \citep{Krajicek-book}). We will provide an example in \Cref{sec:eF_lbs}. In this section, we include the required  background for the result. We assume basic familiarity with model theory.

While the definitions and results presented below can be adapted to other theories of bounded arithmetic, we focus on the theory $\SOT$ for concreteness.

\begin{definition}[Cut in a Model of Arithmetic]
A \emph{cut} in a model $M$ of $\SOT$ is a nonempty set $I \subseteq M$ such that:
\begin{enumerate}
    \item For every $a,b \in M$, if $b \in I$ and $a < b$ then $a \in I$.
    \item For every $a \in M$, if $a \in I$ then $a+ 1 \in I$.
\end{enumerate}
In this case, we write $I \subseteq_e M$.
\end{definition}

Note that a cut is not necessarily closed under operations such as addition and multiplication.

\begin{claim}\label{claim:closed_cut_property}
   Let $M$ be a model of $\SOT$, and let $I \subseteq_e M$. Moreover, assume that $I$ is closed under $+$, $\cdot$, and $\#$ operations. Let $\varphi(a,\vec{b})$ be a bounded formula with all free variables displayed. Let $\vec{v}$ be elements of $I$. Then for every $u \in I$,
   $$
I \models \varphi(u,\vec{v}) \quad \Longleftrightarrow \quad M \models \varphi(u,\vec{v}).
   $$
\end{claim}

\Cref{claim:closed_cut_property} can be proved by induction on the complexity of $\varphi$. Using the claim, one can establish the following lemma.

\begin{lemma}\label{l:closed_cuts_are_models}
    Let $M$ be a model of $\SOT$, and let $I \subseteq_e M$. Moreover, assume that $I$ is closed under $+$, $\cdot$, and $\#$ operations. Then $I$ is a model of $\SOT$.
\end{lemma}

Since it is not hard to check that a cut $I$ as above satisfies the BASIC axioms of $\SOT$, the proof of \Cref{l:closed_cuts_are_models} essentially amounts to verifying that $I$ satisfies the corresponding induction principle (see, e.g.,~\citep[Lemma 5.1.3]{Krajicek-book} for a similar argument).

For a model $M$, we say that $n\in M$ is a \emph{length} if there is $N \in M$ such that $n = |N|$.

\begin{lemma}\label{lemma:majorizing_length_IDelta}
    Let $M_0$ be a nonstandard countable model of $\SOT$. Then there is a \emph{(}countable\emph{)} cut $M$ of $M_0$ that is a model of $\SOT$ and a length $n \in M$, where $n = |e|$ for some nonstandard $e \in M$, for which the following holds. For every $b \in M$ there is a standard number $k$ such that $M \models |b| \leq n^k$.
\end{lemma}

\begin{proof}
    Let $e \in M_0$ be nonstandard, and let $n \eqdef |e|$. Consider the set 
    $$
I_e \eqdef \{a \in M_0 \mid a \leq t(e)~\text{for some}~\LBuss\text{-term}~t(x)\},
    $$
where we compare elements with respect to the interpretation of the relation symbol $\leq$ in $M_0$. Note that $I_e$ is a cut of $M_0$ and $e \in I_e$. Moreover, it is not hard to check that it is  closed under addition, multiplication, and smash operations. By  \Cref{l:closed_cuts_are_models}, $I_e$ is a model of $\SOT$. Finally, by construction, for every $b \in I_e$ we have $b \leq t(e)$ for some $\LBuss$-term $t$. A simple  induction on the structure of $t$ shows the existence of a standard number $k$ such that $|b| \leq n^k$  in $I_e$.
\end{proof}

Finally, we will need the following definition.

\begin{definition}[Cofinal extension]
    We say that an extension $M'$ of a model $M$ is \emph{cofinal} (or $M$ is cofinal in $M'$) if for every $a \in M'$ there is $b \in M$ such that $a \leq b$ in $M'$. If this is the case, we write $M' \supseteq_\mathsf{cf} M$.
\end{definition}

\section{The Strength of Bounded Arithmetic}

In connection with the fundamental research goal mentioned in \Cref{sec:introduction}, research on the provability of complexity bounds has achieved significant progress on two complementary fronts: the \emph{formalization} of several established results from algorithms and complexity within theories of bounded arithmetic, and the \emph{unprovability} of complexity  bounds in the same theories, often conditional on a computational assumption.

In \Cref{sec:form_algs_complex}, we explore what it means to formalize results from algorithms and complexity theory within the framework of bounded arithmetic, highlighting some of the nuances involved. In \Cref{sec:formula_lb_formalization}, we present some concrete details of the formalization of a formula lower bound in $\PV$.

\subsection{Formalization of Results from Algorithms and Complexity}\label{sec:form_algs_complex}

Several central theorems from mathematics and computer science can be proved in bounded arithmetic. They include results from number theory \citep{Woods1981, PWW88}, graph theory and extremal combinatorics \citep{thesis_KO}, randomized  algorithms and probabilistic arguments  \citep{Jerabek-phd, DBLP:journals/corr/abs-1103-5215, thesis_DTML}, probabilistic checkable proofs \citep{DBLP:journals/corr/Pich14},  circuit lower bounds \citep{DBLP:journals/apal/MullerP20}, expander graphs \citep{DBLP:journals/apal/BussKKK20}, linear algebra \citep{DBLP:journals/jacm/TzameretC21}, Zhuk's CSP algorithm \citep{gaysin2023proofcomplexitycsp,  gaysin2024proofcomplexityuniversalalgebra}, etc.~The reader can find numerous other examples in \citep{cook_nguyen_2010, krajicek_2019,  DBLP:journals/apal/MullerP20} and references therein.

In some cases, the formalization of an existing result in bounded arithmetic is  straightforward, specially once an appropriate framework has been developed (e.g.,~the approximate counting framework of \citep{Jerabek07}, which enables the use of tools from probability theory in $\APC_1$). However, sometimes one needs to discover a new proof whose concepts can be defined in the theory and their associated properties  established using the available inductive axioms   (e.g.,~Razborov's formalization of the Switching Lemma \citep{Razborov-switching-l}).

We provide two instructive examples below. The first is a consequence of the formalization of the PCP Theorem  in $\PV$, while the second concerns different ways of formulating a circuit lower bound statement in bounded arithmetic.

\paragraph{The PCP Theorem in $\PV$.} Pich \citep{DBLP:journals/corr/Pich14} proved the PCP Theorem in $\PV$ by formalizing Dinur's proof \citep{DBLP:journals/jacm/Dinur07}. Exploiting the standard connection between PCPs and hardness of approximation,  Pich's result can be used to show that $\PV$ establishes the NP-hardness of approximating the value of a $k$-SAT instance. This means in particular that, for a suitable $\LPV$-function symbol $f$ obtained from Dinur's argument, $\PV$ proves that $f$ is a gap-inducing reduction from the Boolean Formula Satisfiability Problem to $k$-SAT (for a sufficiently large $k$):   
\begin{eqnarray}
    \PV & \vdash & \forall \varphi\, \Big (\mathsf{Fla}(\varphi) \wedge \exists y\,\mathsf{Sat}(\varphi, y) \rightarrow k\text{-}\mathsf{CNF}(f(\varphi)) \wedge \exists z\,\mathsf{Sat}(f(\varphi),z)\Big ) \nonumber \\
    \PV & \vdash & \forall \varphi\, \Big (\mathsf{Fla}(\varphi) \wedge \forall y\,\neg\mathsf{Sat}(\varphi, y) \rightarrow k\text{-}\mathsf{CNF}(f(\varphi)) \wedge \forall z\,\mathsf{Value}_{\leq 1 - \delta}(f(\varphi),z)\Big ) \nonumber 
\end{eqnarray}
where all the  expressions are quantifier-free $\LPV$-formulas: $\mathsf{Fla}(x)$ checks if $x$ is a valid description of a Boolean formula, $k\text{-}\mathsf{CNF}(x)$ checks if $x$ is a valid description of a $k$-CNF, $\mathsf{Sat}(u,v)$ checks if $v$ is a satisfying assignment for $u$, and $\mathsf{Value}_{\leq 1 - \delta}(u,v)$ holds if $v$ satisfies at most a $(1 - \delta)$-fraction of the clauses in $u$ (with $\delta>0$ being a universal constant from the formalized Dinur's proof).

In the formalization the key point is that $\PV$ proves that the function symbol $f$ behaves as expected. In practice, in order to achieve this, a typical formalization is presented in a semi-formal way, and might claim on a few occasions that some algorithm $f_1$ constructed in a particular way from
another algorithm $f_2$ can be defined in $\PV$. This means that $\PV$ proves that
$f_1$ behaves as described in the definition. This is possible thanks to Cobham's characterization of $\FP$ and the axioms of $\PV$, which ensure that the theory ``understands'' how different algorithms are constructed from one another.  In many cases, the verification that $\PV$ proves the desired properties is straightforward but tedious, requiring some initial setup of basic capabilities  of $\PV$ (often referred to as ``bootstrapping'') which is part of the standard
background in bounded arithmetic.

\paragraph{Circuit Lower Bound Statements.} We discuss two ways of formalizing a complexity lower bound. In this example, for a given size bound $s(n)$ (e.g.,~$s(n) = n^2$), we consider an $\LPV$-sentence $\mathsf{FLB}_{s}^\oplus$  stating that Boolean formulas for the parity function on $n$ bits require at least $s(n)$ leaves:
\begin{equation}
\forall N\, \forall n\,\forall F\, (n = |N| \wedge n \geq 1 \wedge \mathsf{Fla}(F) \wedge  \mathsf{Size}(F) < s(n) \rightarrow \exists x\,(|x| \leq n \wedge \mathsf{Eval}(F,x) \neq \oplus(x))\,, \nonumber    
\end{equation}
where we identify $n$-bit strings with natural numbers of length at most $n$, and employ a well-behaved $\LPV$-function symbol $\oplus$ such that $\PV$ proves the basic properties of the parity function, e.g., $\PV \vdash \oplus(x1) = 1 - \oplus(x)$.\footnote{We often abuse notation and treat $x$ as a string in semi-formal discussions.} 

Note that $\mathsf{FLB}_{s}^\oplus$ is a $\forall \Sigma^b_1$-sentence. Consequently, if $\PV \vdash \mathsf{FLB}_s^\oplus$, we obtain via Herbrand's Theorem (\Cref{thm:Herbrand}) a polynomial-time algorithm $A$ that, when given $N$ of length $n$ and the description of an $n$-bit formula $F$ of size $< s(n)$, $A(N,F)$ outputs a string $x \in \{0,1\}^n$ such that $F(x) \neq \oplus(x)$. In other words, circuit lower bounds provable in $\PV$ are constructive in the sense that they also provide an efficient refuter witnessing that $F$ does not compute parity (see \citep{CJSW21} for more on this topic).

The aforementioned formalization is informally referred to as a ``Log'' formalization of circuit lower bounds. This is because the main parameter $n$ is the length of a variable $N$ and all objects quantified over are of length polynomial in $n$. It is also possible to consider a formalization where $n = ||N||$ ($n$ is the length of the length of $N$), which is known as a ``LogLog'' formalization. This  allows us to quantify over exponentially larger objects, e.g., under such a formalization the entire truth-table of a formula $F$ has length polynomial in the length of $N$. 

Obtaining a Log formalization (e.g.,~\citep{DBLP:journals/apal/MullerP20}) is a stronger result than obtaining a LogLog formalization (e.g.,~\citep{Razborov-switching-l}). In particular, in contrast to the discussion above, a witnessing theorem applied to a LogLog formalization provides a refuter with access to $N$ and thus running in time $\mathsf{poly}(N) = \mathsf{poly}(2^n)$. Conversely, the unprovability of a LogLog circuit lower bound statement (e.g.,~\citep{DBLP:conf/stoc/PichS21, LO23}) is a stronger result than the unprovability of a  Log statement. We refer to the introduction of \citep{DBLP:journals/apal/MullerP20} for a more extensive discussion on this matter.

\subsection{Concrete Example: Subbotovskaya's Formula Lower Bound in \texorpdfstring{$\PV$}{PV}}\label{sec:formula_lb_formalization}

In this section, we explore some details of a formalization in $\PV$ that the parity function $\oplus$ on $n$ bits requires Boolean formulas of size $\geq n^{3/2}$ \citep{subbotovskaya1961realization}. We follow the notation introduced in \Cref{sec:form_algs_complex}. 

\begin{theorem}[\citep{CKKOT24}]\label{thm:FLB_in_PV}
    Let $s(n) \eqdef n^{3/2}$. Then $\PV \vdash \mathsf{FLB}_{s}^\oplus$.
\end{theorem}

The formalization is an adaptation of the argument presented in \citep[Section 6.3]{jukna2012boolean}, which proceeds as follows:
\begin{enumerate}
    \item \citep[Lemma 6.8]{jukna2012boolean}: For any formula $F$ on $n$-bit inputs, it is possible to fix one of its variables so that the resulting formula $F_1$ satisfies $\mathsf{Size}(F_1) \leq (1 - 1/n)^{3/2} \cdot \mathsf{Size}(F)$.
    \item \citep[Theorem 6.10]{jukna2012boolean}: If we apply this result $\ell \eqdef n - k$ times, we obtain a formula $F_{\ell}$ on $k$-bit inputs such that
    $$
\mathsf{Size}(F_\ell) \leq \mathsf{Size}(F) \cdot (1 - 1/n)^{3/2} \cdot (1 - 1/(n-1))^{3/2}  \ldots  (1 - 1/(k+1))^{3/2} = \mathsf{Size}(F) \cdot (k/n)^{3/2}.
    $$
    \item \citep[Example 6.11]{jukna2012boolean}: Finally, if the initial formula $F$ computes the parity function, by setting $\ell = n - 1$ we get
    $
1 \leq \mathsf{Size}(F_\ell) \leq (1/n)^{3/2} \cdot \mathsf{Size}(F),
    $
    and consequently $\mathsf{Size}(F) \geq n^{3/2}$.
\end{enumerate}

We present the argument in a more constructive way when formalizing the result in $\PV$. In more detail, given a small formula $F$, we recursively construct (and establish correctness by induction) an $n$-bit input $y$ witnessing that $F$ does not compute the parity function.\footnote{Actuallly, for technical reasons related to the induction step, we will simultaneously construct an $n$-bit input $y^0_n$ witnessing that $F$ does not compute the parity function and an $n$-bit input $y^1_n$ witnessing that $F$ does not compute the negation of the parity function.}

\begin{proof} We follow closely the presentation from \citep{CKKOT24}. For brevity, we only discuss the formalization of the main inductive argument. More details can be found in \citep{CKKOT24}. Given $b \in \{0,1\}$, we introduce the function $\oplus^b(x) \eqdef \oplus(x) + b\;(\mathsf{mod}\;2)$.
   In order to prove $\mathsf{FLB}_{s}^\oplus$ in $\PV$, we explicitly consider a polynomial-time function $R(1^n, F,b)$ with the following property:\footnote{For convenience, we often write $1^n$ instead of explicitly considering parameters $N$ and $n = |N|$. In practice, it means that $R$ gets as input $N$ (together with other parameters) but with respect to $N$ it only depends on $n = |N|$.}
   \begin{center}
   If $\mathsf{Size}(F) < s(n)$ then $R(1^n,F,b)$ outputs an $n$-bit string $y^b_n$ such that $\mathsf{Eval}(F,y^b_n) \neq \oplus^b(y^b_n)$.
   \end{center}
   In other words, $R(1^n,F,b)$ witnesses that the formula $F$ does not compute the function $\oplus^b$ over $n$-bit strings. Note that the correctness of $R$ is captured by a sentence $\mathsf{Ref}_{R,s}$ described as follows:\footnote{Similarly, the notation $\forall 1^n$ denotes $\forall N \forall n$ but we add the condition that $n = |N|$ in the subsequent formula. We might also write just $F(x)$ instead of $\mathsf{Eval}(F,x)$}
   \begin{equation}
   \forall 1^n\, \forall F\,(\mathsf{Fla}(F) \wedge \mathsf{Size}(F) < s(n) \rightarrow |y^0_n|_\ell = |y^1_n|_\ell = n \wedge  F(y^0_n) \neq \oplus^0(y^0_n) \wedge F(y^1_n) \neq \oplus^1(y^1_n))\,, 
       \nonumber
   \end{equation}
   where we employ the abbreviations $y^0_n \eqdef R(1^n,F,0)$ and $y^1_n \eqdef  R(1^n,F,1)$, and for convenience use $|z|_\ell$ to denote the bitlength of $z$. 
   Our plan is to define $R$ and show that $\PV \vdash \mathsf{Ref}_{R,s}$. Note that this implies $\mathsf{FLB}_{s}^\oplus$ in $\PV$ by standard first-order logic reasoning. 
   
   The correctness of $R(1^n, F,b)$ will be established by polynomial induction on $N$ (equivalently, induction on $n = |N|$). Since $\mathsf{Ref}_{R,s}$ is a universal sentence and $\SOT(\LPV)$ is $\forall \Sigma^b_1$-\emph{conservative} over $\PV$ (i.e., provability of such a sentence in $\SOT(\LPV)$ implies its provability in $\PV$), it is sufficient to describe a formalization in the more convenient theory $\SOT(\LPV)$. For this reason, polynomial induction for $\NP$ and $\coNP$ predicates (admissible in $\SOT(\LPV)$; see, e.g.,~\citep[Section 5.2]{Krajicek-book}) is available during the formalization. More details follow.

The procedure $R(1^n,F,b)$ makes use of a few polynomial-time sub-routines (briefly discussed in the comments in the pseudocode below) and is defined in the following way:\\

   \begin{algorithm}[H]
     \SetKwInOut{Input}{Input}
     \caption{Refuter Algorithm $R(1^n,F,b)$  \citep{CKKOT24}.}\label{algo:RefuterR}
     \Input{$1^n$ for some $n \geq 1$, formula $F$ over $n$-bit inputs, $b \in \{0,1\}$.}
     Let $s(n) \eqdef n^{3/2}$. If $\mathsf{Size}(F) \geq s(n)$ or $\neg \mathsf{Fla}(F)$  \Return{``error''}\;
     If $\mathsf{Size}(F) = 0$, $F$ computes a constant function $b_F \in \{0,1\}$. In this case, \Return{the $n$-bit string $y^b_n \eqdef y^b_1 0^{n-1}$ such that $\oplus^b(y^b_1 0^{n-1}) \neq b_F$}\;
     Let $\widetilde{F} \eqdef \mathsf{Normalize}(1^n, F)$\; \tcp{$\widetilde{F}$ satisfies the  conditions in the proof of \citep[Claim 6.9]{jukna2012boolean}, $\mathsf{Size}(\widetilde{F}) \leq \mathsf{Size}(F)$, $\forall x \in \{0,1\}^n\;F(x) = \widetilde{F}(x)$.}
     Let $\rho \eqdef \mathsf{Find}\text{-}\mathsf{Restriction}(1^n,\widetilde{F})$, where $\rho \colon [n] \to \{0,1,\star\}$ and $|\rho^{-1}(\star)| = n-1$\; \tcp{$\rho$ restricts a suitable variable $x_i$ to a bit $c_i$, as in \citep[Lemma 6.8]{jukna2012boolean}.}
     Let $F' \eqdef \mathsf{Apply}\text{-}\mathsf{Restriction}(1^n,\widetilde{F}, \rho)$. Moreover, let $b' \eqdef b \oplus c_i$ and $n' \eqdef n -1$\;
\tcp{$F'$ is an $n'$-bit formula; $\forall z \in \{0,1\}^{\rho^{-1}(\star)}\;F'(z) =  \widetilde{F}(z \cup x_i \mapsto c_i)$.}
     Let  $y^{b'}_{n'} \eqdef R(1^{n'}, F', b')$ and \Return{the $n$-bit string $y^b_{n} \eqdef y^{b'}_{n'} \cup y_i \mapsto c_i$}\;
   \end{algorithm}

   \vspace{0.2cm}

\noindent (The pseudocode presented above is only an informal specification of $R(1^n,F,b)$. As mentioned in \Cref{sec:form_algs_complex}, a completely formal proof in $\PV$ would employ Cobham's formalism and would specify how $R(1^n,F,b)$ can be defined from previously defined algorithms (e.g.,~$\mathsf{Apply}\text{-}\mathsf{Restriction}$) via the allowed operations.)

We note that $R(1^n,F,b)$ runs in time polynomial in $n + |F| + |b|$ and that it is definable in $\SOT(\LPV)$. Next, as an instructive example, we establish the correctness $R(1^n,F,b)$ in $\SOT(\LPV)$ by \emph{polynomial induction} (PIND) for $\Pi^b_1$-formulas, assuming that the subroutines appearing in the pseudocode of $R(1^n,F,b)$ satisfy the necessary properties (provably in $\SOT(\LPV)$).

 \begin{lemma} Let $s(n) \eqdef n^{3/2}$. Then $\SOT(\LPV) \vdash \mathsf{Ref}_{R,s}$.
 \end{lemma}
\begin{proof}
    We consider the formula $\varphi(N)$ defined as 
    $$
  \forall F\,\forall n\,(n = |N| \wedge n \geq 1 \wedge \mathsf{Fla}(F) \wedge \mathsf{Size}(F) < s(n)) \rightarrow \quad \quad \quad \quad \quad \quad \quad \quad \quad \quad
  $$
  $$\quad \quad \quad \quad \quad \quad \quad \quad \quad \quad (|y^0_n|_\ell = |y^1_n|_\ell = n \wedge  F(y^0_n) \neq \oplus^0(y^0_n) \wedge F(y^1_n) \neq \oplus^1(y^1_n))\,, 
   $$
   where as before we use $y^0_n \eqdef R(1^n,F,0)$ and $y^1_n \eqdef R(1^n,F,1)$. Note that $\varphi(N)$ is a $\Pi^b_1$-formula. Below, we argue that
   $$
\SOT(\LPV) \vdash \varphi(1) \quad \text{and} \quad \SOT(\LPV) \vdash \forall N\,\varphi(\lfloor N/2 \rfloor) \to \varphi(N)\,.
   $$
   Then, by polynomial induction for $\Pi^b_1$-formulas (available in $\SOT(\LPV)$) and using that $\varphi(0)$ trivially holds, it follows that $\SOT(\LPV) \vdash \forall N\,\varphi(N)$. In turn, this yields $\SOT(\LPV) \vdash \mathsf{Ref}_{R,s}$.

   \paragraph{Base Case: $\SOT(\LPV) \vdash \varphi(1)\,$.} In this case, for a given formula $F$ and length $n$, the hypothesis of $\varphi(1)$ is satisfied only if $n = 1$, $F$ is a valid description of a formula, and $\mathsf{Size}(F) = 0$. Let $y^0_1 \eqdef R(1,F,0)$ and $y^1_1 \eqdef R(1,F,1)$. We need to prove that
   $$
|y^0_1|_\ell = |y^1_1|_\ell = 1 \wedge F(y^0_1) \neq \oplus^0(y^0_1) \wedge F(y^1_1) \neq \oplus^1(y^1_1)\,. 
   $$
Since $n=1$ and $\mathsf{Size}(F) = 0$, $F$ evaluates to a constant $b_F$ on every input bit. The statement above is implied by Line $2$ in the definition of $R(n,F,b)$. 

   \paragraph{(Polynomial) Induction Step: $\SOT(\LPV) \vdash \forall N\,\varphi(\lfloor N/2 \rfloor) \to \varphi(N)\,$.} Fix an arbitrary $N$, let $n \eqdef |N|$, and assume that $\varphi(\lfloor N/2 \rfloor)$ holds. By the induction hypothesis, for every valid formula $F'$ with $\mathsf{Size}(F') < n'^{3/2}$, where $n' \eqdef n - 1$, we have
   \begin{equation}\label{eq:ind_hyp_final}
   |y^0_{n'}|_\ell = |y^1_{n'}|_\ell = n' \;\wedge\; F'(y^0_{n'}) \neq \oplus^0(y^0_{n'}) \;\wedge\; F'(y^1_{n'}) \neq \oplus^1(y^1_{n'})\,,
   \end{equation}
where $y^0_{n'} \eqdef R(1^{n'}, F', 0)$ and $y^1_{n'} \eqdef R(1^{n'}, F', 1)$. 

Now let $n \geq 2$, and let $F$ be a valid description of a formula over $n$-bit inputs with $\mathsf{Size}(F) < n^{3/2}$. By the size bound on $F$, $R(1^n,F,b)$ ignores Line 1. If $\mathsf{Size}(F) = 0$, then similarly to the base case it is trivial to check that the conclusion of $\varphi(N)$ holds. Therefore, we assume that $\mathsf{Size}(F) \geq 1$ and $R(1^n,F,b)$ does not stop at Line 2. 

Consider the following definitions: 

\begin{multicols}{2}
\begin{enumerate}
    \item $\widetilde{F} \eqdef \mathsf{Normalize}(1^n, F)$ (Line 3),
    \item $\rho \eqdef \mathsf{Find}\text{-}\mathsf{Restriction}(1^n,\widetilde{F})$ (Line 4),
    \item $F' \eqdef \mathsf{Apply}\text{-}\mathsf{Restriction}(1^n,\widetilde{F},\rho)$ (Line 5),
    \item $n' \eqdef n - 1$ (Line 5),
    \item $b' \eqdef b \oplus c_i$ (Line 5), where $\rho$ restricts $x_i$ to $c_i$,
    \item $y^{b'}_{n'} \eqdef R(1^{n'}, F', b')$ (Line 6),
    \item $y^b_n \eqdef y^{b'}_{n'} \cup y_i \mapsto c_i$ (Line 6),
    \item $s \eqdef \mathsf{Size}(F)$, $\widetilde{s} \eqdef \mathsf{Size}(\widetilde{F})$, and $s' \eqdef \mathsf{Size}(F')$. 
\end{enumerate}
\end{multicols}

\noindent We rely on the provability in $\SOT(\LPV)$ of the following statements about  the subroutines of $R(1^n,F,b)$ (see \citep{CKKOT24}):
\begin{multicols}{2}
\begin{itemize}
    \item[(\emph{i})] $\widetilde{s} \leq s$\,,
    \item[(\emph{ii})] $s' \leq \widetilde{s} \cdot (1 - 1/n)^{3/2}$\,,
    \item[(\emph{iii})] $\forall x \in \{0,1\}^n\;\widetilde{F}(x) = F(x)$\,,
    \item[(\emph{iv})] $\forall z \in \{0,1\}^{\rho^{-1}(\star)}\;F'(z) =  \widetilde{F}(z \cup x_i \mapsto c_i)$\,.
\end{itemize}
\end{multicols}
\noindent By Items (\emph{i}) and (\emph{ii}) together with the bound $s < n^{3/2}$,
$$
\SOT(\LPV) \vdash s' \leq \widetilde{s} \cdot (1 - 1/n)^{3/2} \leq s \cdot (1 - 1/n)^{3/2} < n^{3/2} \cdot (1 - 1/n)^{3/2} = (n - 1)^{3/2}\,.
$$
Thus $F'$ is a valid formula on $n'$-bit inputs of size $< n'^{3/2}$. By the first condition in the induction hypothesis (\Cref{eq:ind_hyp_final}) and the definition of each $y^b_n$, we have $|y^0_n|_\ell = |y^1_n|_\ell =  n$. Using the definitions listed above, the last two conditions in the induction hypothesis (\Cref{eq:ind_hyp_final}), and Items (\emph{iii}) and (\emph{iv}), we derive in $\SOT(\LPV)$ the following statements for each $b \in \{0,1\}$:
\begin{eqnarray}
        F'(y^{b'}_{n'}) & \neq & \oplus^{b'}(y^{b'}_{n'})\,, \nonumber \\
        F(y^b_n) & = & F'(y^{b'}_{n'})\,, \nonumber  \\
        F(y^b_n) & \neq & \oplus^{b'}(y^{b'}_{n'})\,. \nonumber  
\end{eqnarray}
Therefore, using basic facts about the function symbols $\oplus^0$ and $\oplus^1$,
$$
        \oplus^{b'}(y^{b'}_{n'})  =  \oplus^{b \oplus c_i}(y^{b'}_{n'})  =  c_i \oplus (\oplus^b(y_{n'}^{b'}))  =   c_i \oplus (\oplus^b(y_{n}^{b}) \oplus c_i) = \oplus^b(y^b_n)\,.
$$
These statements imply that, for each $b \in \{0,1\}$, $F(y^b_n) \neq \oplus^b(y^b_n)$. In other words, the conclusion of $\varphi(N)$ holds. This completes the proof of the induction step.
\end{proof}

 As explained above, the provability of $\mathsf{Ref}_{R,s}$ in $\SOT(\LPV)$ implies its provability in $\PV$. Since $\PV \vdash \mathsf{Ref}_{R,s} \rightarrow \mathsf{FLB}_{s}^\oplus$, this completes the proof of \Cref{thm:FLB_in_PV}.
\end{proof}

We have seen that a non-trivial formula size  lower bound can be established in $\PV$. More advanced circuit lower bounds are known to be provable assuming additional axioms extending $\PV$ (e.g.,~\citep[Section 15.2]{Krajicek-book} and \citep{DBLP:journals/apal/MullerP20}), but their provability in $\PV$ (or equivalently, in $\SOT(\LPV)$) is less clear.

\begin{problem}
 For each $d \geq 1$ and $\ell \geq 1$, can $\PV$ prove that the parity function on $n$ bits cannot be computed by depth-$d$ circuits of size $n^\ell$?
 \end{problem}

 \begin{problem} For each $\ell \geq 1$, is there a constant $k = k(\ell)$ such that $\PV$ proves that every monotone circuit for the $k$-clique problem on $n$-vertex graphs  must be of size at least $n^\ell$?
\end{problem}

\section{Unprovability of Complexity Bounds}\label{sec:unprov}

The investigation of the unprovability of complexity bounds within theories of bounded arithmetic has a long and rich history. Much of the early work  took place in the nineties, with significant  results obtained by Razborov \citep{Razborov-switching-l, razborov1995unprovability}, Kraj{\'{i}}{\v{c}}ek \citep{DBLP:journals/jsyml/Krajicek97}, and other researchers. Since then, and in particular over the last decade, there has been renewed interest and  progress in establishing unprovability results (see, e.g., \citep{DBLP:journals/jsyml/CookK07, DBLP:conf/stoc/PichS21, CKKO21, LO23, DBLP:conf/stoc/AtseriasBM23} and references therein).

In \Cref{sec:unprov_upper_bounds}, we consider the unprovability of complexity upper bounds. The \emph{unprovability} of an inclusion such as $\mathsf{NP} \subseteq \mathsf{SIZE}[n^k]$ is equivalent to the \emph{consistency} of $\mathsf{NP} \nsubseteq \mathsf{SIZE}[n^k]$ with the corresponding theory. Such a consistency result establishes that, while we cannot confirm the separation is true in the standard model of natural numbers, we know it holds in a non-standard model of a theory so strong that complexity theory appears almost indistinguishable from the standard one. We stress that establishing the consistency of a lower bound is a necessary step towards showing that the lower bound is true. For this reason, the unprovability of  upper bounds can be formally seen as progress towards showing unconditional complexity lower bounds.

In \Cref{sec:unprov_lower_bounds}, we turn our attention to the unprovability of complexity lower bounds. This direction is partly driven by the desire to formally understand why proving complexity lower bounds is challenging, and to explore the possibility of a more fundamental underlying reason for this difficulty. Moreover, it might provide 
 examples of hard sentences for logical theories and of hard propositional tautologies for proof systems. The investigation of the meta-mathematics of lower bounds has also found unexpected applications in algorithms and complexity (e.g.,~\citep{DBLP:conf/coco/CarmosinoIKK16}).

Finally, in \Cref{sec:connection_lbs_ubs} we connect the two directions and explain how the unprovability of circuit lower bounds in $\PV$ yields the unprovability of $\mathsf{P} = \mathsf{NP}$ in $\PV$. The latter can be seen as a weakening of the $\P$ versus $\NP$ problem that considers the existence of feasible proofs that $\P = \NP$. This further motivates the investigation of the unprovability of lower bounds.

\subsection{Unprovability of Upper Bounds}\label{sec:unprov_upper_bounds}

\subsubsection{LEARN-Uniform Circuits and Unprovability}\label{sec:LEARN_unprov}

Cook and Kraj{\'{i}}{\v{c}}ek \citep{DBLP:journals/jsyml/CookK07} considered the provability of $\mathsf{NP} \subseteq \mathsf{SIZE}[\mathsf{poly}]$ in bounded arithmetic and obtained a number of conditional negative results. \cite{KO17}, building on techniques from \citep{DBLP:journals/jsyml/CookK07}, showed that for no integer $k \geq 1$ the theory $\PV$ proves that  $\mathsf{P} \subseteq \mathsf{SIZE}[n^k]$. Note that this is an unconditional result. Thus, for a natural theory capable of formalizing advanced results from complexity theory, such as the PCP Theorem, we can unconditionally rule out the provability of $\mathsf{P} \subseteq \mathsf{SIZE}[n^k]$. A slightly stronger model-theoretic formulation of the result of \cite{KO17} appears in \citep{DBLP:journals/aml/BydzovskyM20}.

\citep{BKO20} obtained results for stronger theories and ruled out the provability of infinitely often inclusions. In more detail, for an $\LPV$-function symbol $h$, consider the sentence 
\begin{equation}
\upci[h] \eqdef \forall 1^m\,\exists 1^n\, \exists C_n\, \forall x\, \big (n \geq m \wedge |C_n| \leq n^k \wedge (|x|\leq n \rightarrow \psi(n,C_n,x,h)) \big)\,,\footnote{Recall that $1^n$ is simply a convenient notation to refer to a variable $n$ that is set to $|N|$ for some variable $N$.}
    \nonumber
\end{equation}
where $\psi$ is a quantifier-free $\LPV$-formula stating that $h(x) \neq 0$ if and only if the evaluation of the circuit $C_n$ on $x$ (viewed as an $n$-bit string) is $1$. In other words, $\upci[h]$ states that the language defined by $h$ (which is in $\mathsf{P}$) admits circuits of size at most $n^k$ on infinitely many input lengths $n$. \citep{BKO20} showed that for each $k \geq 1$, there is an $\LPV$-function symbol $h$ such that $\PV$ does not prove $\upci[h]$. Similarly, they established that $\SOT \nvdash \mathsf{NP} \subseteq \mathsf{i.o.}\mathsf{SIZE}[n^k]$ and $\TOT \nvdash \mathsf{P}^{\mathsf{NP}} \subseteq \mathsf{i.o.}\mathsf{SIZE}[n^k]$. 

Building on these results, \citep{CKKO21} introduced a modular framework to establish the unprovability of circuit upper bounds in bounded arithmetic using a learning-theoretic perspective. Next, we describe how their approach can be used to show a slightly weaker form of the result from \citep{BKO20} described above. For an $\LPV$-function symbol $h$, we consider a sentence $\mathsf{UB}_{c,k}[h]$ stating that $L_h \in \mathsf{SIZE}[c \cdot n^k]$, where $x \in L_h$ if and only if $h(x) \neq 0$, i.e.,
\begin{equation}\label{eq:ckh}
\mathsf{UB}_{c,k}[h] \eqdef \forall 1^n\,\exists C_n\,\forall x\, \big ( |C_n| \leq c \cdot n^k \wedge (|x| \leq n \rightarrow (\mathsf{Eval}(C_n,x,n) = 1 \leftrightarrow h(x) \neq 0)) \big )\,, 
\end{equation}
where $\mathsf{Eval}(C_n,x,n)$ is an $\LPV$-function that evaluates the circuit $C_n$ on the $n$-bit string described by $x$.  Our goal is to show that for every $k \geq 1$ there is a function symbol $h$ such that, for no choice of $c \geq 1$, $\PV$ proves $\mathsf{UB}_{c,k}[h]$. (Note that in all results discussed in this section, we consider Log formalizations, as explained in \Cref{sec:form_algs_complex}.)\\ 

\noindent \textbf{Overview of the Approach.} Note that $\mathsf{UB}_{c,k}[h]$ claims the  \emph{existence} of circuits for $L_h$, i.e., it states a \emph{non-uniform} upper bound. We explore the constructive aspect of $\PV$ proofs, by extracting computational information from a $\PV$-proof that such circuits exist. The argument has a \emph{logical component}, where we extract from a proof of $\mathsf{UB}_{c,k}[h]$ a ``LEARN-uniform'' construction of a sequence $\{C_n\}_n$ of circuits for $L_h$, and a \emph{complexity-theoretic component}, where we unconditionally establish that for each $k$ LEARN-uniform circuits of this form do not exist for some $h$. Altogether, we get that for some $h$ theory $\PV$ does not prove $\mathsf{UB}_{c,k}[h]$ (no matter the choice of $c$).\\

\noindent \textbf{LEARN-uniform circuits.} We will be interested in languages that can be efficiently learned with a bounded number of equivalence queries, in the following sense. For functions $s,q \colon \mathbb{N} \to \mathbb{N}$, we say that a language $L \subseteq \{0,1\}^*$ is in $\mathsf{LEARN}\text{-}\mathsf{uniform}^{\mathsf{EQ}[q]}~\mathsf{SIZE}[s]$ if there is a polynomial-time algorithm $A^{\mathsf{EQ}(L_n)}(1^n)$ that outputs a circuit of size at most $s(n)$ for $L_n$ after making at most $q(n)$ equivalence queries to $L_n$, where $L_n = L \cap \{0,1\}^n$. The equivalence query oracle, given the description of an $n$-bit circuit $D$ of size a most $s(n)$, replies ``yes'' if $D$ computes $L_n$, or provides some counter-example $w$ such that $D(w) \neq L_n(w)$.\\

\noindent \textbf{Extracting LEARN-uniform circuits from $\PV$ proofs.} For convenience, write $\mathsf{UB}_{c,k}[h] = \forall 1^n\,\exists C_n\,\forall x\;\phi(1^n,C_n,x)$ in \Cref{eq:ckh}, where $\phi(1^n,C_n,x)$ is a quantifier-free formula. Since $\PV$ is a universal theory, under the assumption that $\PV \vdash \mathsf{UB}_{c,k}[h]$, we can apply \Cref{thm:KPT} (KPT Witnessing Theorem) to obtain the provability in $\PV$ of the disjunction
\begin{equation}
\forall 1^n \;\forall x_1  \ldots \forall x_k \;
    \Big ( \phi(1^n,t_1(1^n),x_1) \lor \phi(1^n,t_2(1^n,x_1),x_2) 
    \lor \dots \lor \phi(1^n,t_k(1^n,x_1, \dots ,x_{k-1}), x_k)\Big ), \label{eq:KPT_game}
\end{equation}
where $t_1, \ldots, t_k$ are $\LPV$-terms and $k = O(1)$. Most importantly, due to the soundness of $\PV$, this statement is true over the standard model $\mathbb{N}$. Additionally, the terms in $\PV$ correspond to polynomial-time algorithms. Next, we will discuss how to interpret \Cref{eq:KPT_game} over $\mathbb{N}$ as an interactive protocol and how this perspective leads to a LEARN-uniform construction.

The KPT Witnessing Theorem can be intuitively understood as follows \citep{KrajicekPS90}. Consider a search problem $Q(1^n)$, where given the input $1^n$, we need to find $D$ such that $\forall x\,\phi(1^n,D,x)$. The problem $Q(1^n)$ can be solved using a $k$-round \emph{Student-Teacher protocol}.
 In the first round, the student proposes $D_1 = t_1(1^n)$ as a solution to the search problem $Q(1^n)$. This solution is either correct, or there exists a counterexample $w_1$ such that $\neg \phi(1^n,t_1(1^n),w_1)$. The teacher then provides this counterexample value $w_1$, and the protocol moves to the next round. In each subsequent round $1 \leq i < k$, the student computes $D_i = t_i(1^n,w_1,\dots,w_{i-1})$ based on the counterexamples $w_1,\dots,w_{i-1}$ received in the previous rounds. This $D_i$ is either a correct solution for $Q(1^n)$, in which case the problem is solved, or there is another counterexample $w_i$ provided by the teacher such that $\neg \phi(1^n, t_i(1^n,w_1,\dots,w_{i-1}),w_i)$. If the latter is the case, the protocol continues to the next round $i+1$.
The theorem guarantees that for every input $1^n$, the student will successfully solve the search problem $Q(1^n)$ within some round $1 \leq i \leq k$. 

From a $\PV$ proof of a circuit upper bound for a language $L_h$, we can derive a Student-Teacher protocol for the search problem $Q(1^n)$ corresponding to \Cref{eq:KPT_game}. In this protocol, the student proposes a candidate circuit $D$, and the teacher provides a counterexample $w$ to $D$ (an input $w$ such that $D(w) \neq L_h(w)$) if one exists. (Note that $\phi(1^n,D,x)$ might not be true for other reasons, e.g., if $|D| > c \cdot n^k$, but in such cases there is no need to invoke the equivalence query oracle and we can proceed in the Student-Teacher protocol with, say, $w = 0^n$.) The student is guaranteed to succeed after at most $k$ queries, regardless of the counterexamples provided by the teacher.  Finally, for every input $n$, the student computes according to a constant number of fixed $\PV$ terms $t_1, \ldots, t_k$. Since a $\PV$ term is merely a composition of a finite number of $\PV$ function symbols (polynomial-time algorithms), the student's computation runs in  polynomial time. Therefore, from the provability in $\PV$ of a non-uniform circuit upper bound for a language  $L \in \P$, we can extract a LEARN-uniform family of circuits for $L$.\\

\noindent \textbf{Unconditional lower bound against LEARN-uniform circuits.} The argument described above reduces the unprovability of upper bounds to a complexity-theoretic question with no reference to logic. To complete the proof, it is enough to show that for each $k$ there is a language $L \in \P$ such that $L \notin \mathsf{LEARN}\text{-}\mathsf{uniform}^{\mathsf{EQ}[O(1)]}~\mathsf{SIZE}[O(n^k)]$. This unconditional lower bound against LEARN-uniform circuits is established in \citep{CKKO21} by generalizing a lower bound from
\citep{DBLP:journals/cc/SanthanamW14} against $\P\text{-}\mathsf{uniform}$ circuits, which can be interpreted as LEARN-uniform constructions with $q = 0$ queries. Roughly speaking, \citep{CKKO21} shows that one can eliminate each equivalence query using a small amount of non-uniform advice, and that the base case where no queries are present (as in \citep{DBLP:journals/cc/SanthanamW14}) can be extended to a lower bound against a bounded amount of advice.\\

This completes the sketch of the argument. The approach is fairly general and  can be adapted to other theories. The strength of the theory affects the learning model against which one needs to obtain lower bounds (e.g.,~by increasing the number of queries or allowing randomized learners).

\begin{problem}\label{problem:S12_P}
    Show that $\SOT$ does not prove that $\P \subseteq \mathsf{SIZE}[n^k]$.
\end{problem}

In order to solve \Cref{problem:S12_P}, using the connection from \citep{CKKO21} it is sufficient to show that $\P \nsubseteq \mathsf{LEARN}\text{-}\mathsf{uniform}^{\mathsf{EQ}[q]}~\mathsf{SIZE}[O(n^k)]$ for $q = \mathsf{poly}(n)$. In other words, this amounts to understanding the class of languages that admit circuits that can be produced with a polynomial number of equivalence queries.

\begin{problem}
    Show that $\TOT$ does not prove that $\NP \subseteq \mathsf{SIZE}[n^k]$.
\end{problem}

\subsubsection{$\mathsf{P} = \mathsf{NP}$ and  Propositional Proof Complexity}

Suppose that $\mathsf{P}$ is actually equal to $\mathsf{NP}$. In this scenario, there exists a polynomial-time algorithm $g$ (i.e., a $\PV$ function symbol) that can find a satisfying assignment for any given satisfiable formula. In other words, if $\mathsf{Formula}(F,1^n)$ denotes an $\LPV$-formula that checks if $F$ is a valid description of a formula over $n$ input bits, and $\mathsf{Sat}(F,x)$ is an $\LPV$-formula 
that checks if $x$ satisfies the formula encoded by $F$, the sentence
\begin{equation} \label{31.1.19a}
\pnp[g]  \;\eqdef\; \forall 1^n\,\forall F\, \forall x\,\big ( (\mathsf{Formula}(F,1^n) \wedge \mathsf{Sat}(F,x)) \rightarrow \mathsf{Sat}(F, g(F)) \big )
\end{equation}
is true in the standard model $\mathbb{N}$.

\begin{problem} \label{open:PV_proves_PNP}
Show that for no polynomial-time function symbol $g$  theory $\PV$ proves the sentence $\pnp[g]$.
\end{problem}

Equivalently, \Cref{open:PV_proves_PNP} states that $\PV$ (and by standard conservation results  $\mathsf{S}^1_2$) is consistent with $\mathsf{P} \neq \mathsf{NP}$.
This means that either $\mathsf{P} \neq \mathsf{NP}$, as is commonly assumed, making the conjecture trivially true, or $\mathsf{P} = \mathsf{NP}$, but this cannot be proven using only polynomial-time concepts and reasoning. Therefore, \Cref{open:PV_proves_PNP} represents a formal weakening of the conjecture that $\mathsf{P} \neq \mathsf{NP}$. The statement is known to follow from the purely combinatorial conjecture that the extended Frege propositional proof system $\EF$ (see \Cref{sec:prelim_prop_proofs}) is not polynomially bounded, which is a major open problem in proof complexity. 

\begin{theorem}[\citep{Coo75}]\label{thm:PNPEF}
    Suppose that there is a sequence $\{F_n\}_{n \geq 1}$ of propositional tautologies of size polynomial in $n$ that require $\EF$ proofs of size $n^{\omega(1)}$. Then there is no function symbol $g$ such that $\PV$ proves $\pnp[g]$.
\end{theorem}

\begin{proof} Here we only provide a sketch of the proof. More details and extensions of the result can be found in the textbooks \citep{Krajicek-book, krajicek_2019}. We establish  that if $\PV \vdash \pnp[g]$ for some $g$, then every tautology has a polynomial size $\EF$ proof. 

Recall the definitions and results from \Cref{sec:prelim_prop_proofs}. For a propositional proof system $P$ (described by an $\LPV$ function symbol), we consider an $\LPV$-sentence stating the soundness of $P$:
$$
\mathsf{Sound}_P \eqdef \forall 1^n\,\forall F\,\forall \pi\,(\mathsf{Formula}(F,1^n) \wedge \mathsf{Proof}_P(F,\pi)) \rightarrow \forall x\,(|x|\leq n \rightarrow \mathsf{Sat}(F,x))\,,
$$
where $\mathsf{Proof}_P(F,\pi)$ states that $\pi$ is a valid $P$-proof of $F$. 

Note that if $F$ is not a tautology then $g(\neg F)$ outputs a satisfying assignment of $\neg F$, while if $F$ is a tautology then $\neg F$ admits no satisfying assignment. We consider a proof system $P_g$ defined as follows: Given a valid description of an $n$-bit propositional formula $F$ and a candidate proof $\widetilde{\pi}$, $P_g$ accepts $\widetilde{\pi}$ as a proof of $F$ if and only if 
$$
g(\neg F) = \widetilde{\pi} \quad \text{and} \quad \neg \mathsf{Sat}(\neg F, \widetilde{\pi})\,,
$$
where $\neg F$ represents the negation of $F$. Observe that for any tautology $F$, $\pi_F \eqdef g(\neg F)$ is a valid $P_g$-proof of $F$.  

Note that $\PV \vdash \mathsf{Sound}_{P_g}$, which follows from the provability of \Cref{31.1.19a} and the definition of $P_g$ using $g$. Now consider the quantifier-free $\LPV$-formula 
$$
\psi \eqdef \neg \mathsf{Formula}(F,1^n) \vee \neg \mathsf{Proof}_{P_g}(F,\pi) \vee |x| > n \vee \mathsf{Sat}(F,x).
$$
The provability of $\forall 1^n\,\forall F\,\forall \pi\,\psi$ in $\PV$ follows from the provability of $\mathsf{Sound}_{P_g}$. 

Using Cook's translation (\Cref{sec:prelim_prop_proofs}), the sequence of propositional formulas $||\psi||_{m}$ admits $\EF$-proofs of polynomial size. Moreover, given an actual $n$-bit propositional formula $F$ of polynomial size and the corresponding $P_g$-proof $\pi_F$ (represented by fixed strings $\langle F \rangle$ and $\langle \pi_F \rangle$), one can show that there are polynomial size $\EF$ proofs of both $||\mathsf{Formula}(\langle F \rangle, 1^n)||_{\mathsf{poly}(n)}$ and $||\mathsf{Proof}_{P_g}(\langle F \rangle, \langle \pi_F \rangle)||_{\mathsf{poly}(n)}$. (Intuitively, this follows by an evaluation of the expressions on these fixed inputs.) Since $\EF$ is closed under substitution, we can derive in $\EF$ with a polynomial size proof the formula $||\mathsf{Sat}(\langle F \rangle, x)||_{\mathsf{poly}(n)}$.

Finally, for every propositional formula $F(x)$ on $n$-bit inputs, it is possible to efficiently prove in $\EF$ the propositional formula $||\mathsf{Sat}(\langle F \rangle, x)||_{\mathsf{poly}(n)} \rightarrow F(x)$. (This can be established by a slightly more general structural induction on formulas $F$ using information about $|| \cdot ||$ and $\langle \cdot \rangle$.)
Overall, since $\EF$ is closed under implication, it follows from these derivations that there is a polynomial size $\EF$ proof of $F$. This completes the sketch of the proof of the result.
\end{proof}

\Cref{open:PV_proves_PNP} would also follow from a proof that Buss's hierarchy of theories $\mathsf{T}^i_2$ does not collapse \citep{KrajicekPT91}, another central problem in bounded arithmetic. More precisely, it is enough to obtain the following separation.

\begin{problem} Show that for some $i > j \geq 1$ we have $\mathsf{T}^i_2 \neq \mathsf{T}^j_2$.
\end{problem}

It is known that $\PV$ proves that $\P = \NP$ if and only if it proves that $\NP = \coNP$.\footnote{Due to space constraints, we do not elaborate on the formalization of $\NP = \coNP$.} Consequently, a super-polynomial lower bound on the length of $\EF$ proofs also yields the consistency of $\NP \neq \coNP$ with $\PV$.

Finally, we remark that the use of witnessing theorems alone (as done in \Cref{sec:LEARN_unprov}) is probably not sufficient to settle \Cref{open:PV_proves_PNP}. This is because these theorems typically also hold when we extend  the theory with all true universal statements. Thus an unprovability argument that only employs the witnessing theorem would establish unconditionally that each sentence $\pnp[g]$ is false and therefore $\P \neq \NP$. Some researchers interpret this as evidence that the investigation of propositional proof complexity might be unavoidable. Another approach to \Cref{open:PV_proves_PNP} is discussed in \Cref{sec:connection_lbs_ubs}. 

\subsection{Unprovability of  Lower Bounds}\label{sec:unprov_lower_bounds}

\subsubsection{Average-Case Circuit Lower Bounds}

In this section, we discuss the unprovability of strong average-case lower bounds in $\PV$. We focus on an unprovability result from \citep{DBLP:conf/stoc/PichS21}, stated and proved in a slightly stronger form in \citep{LO23}. The proof is based on a technique introduced by \citep{DBLP:journals/jml/Krajicek11} and further explored in \citep{DBLP:journals/apal/Pich15}.

We consider an average-case separation of co-nondeterministic circuits against non-deterministic circuits of subexponential size. In more detail, we investigate the provability of a sentence $\mathsf{LB}^1(s_1,s_2,m,n_0)$ stating that, for every input length $n \geq n_0$, there is a co-nondeterministic circuit $C$ of size $\leq s_1(n)$ such that, for every nondeterministic circuit $D$ of size $\leq s_2(n)$, we have
$$
\Pr_{x \sim \{0,1\}^n}\Big [C(x) = D(x) \Big] \;\leq\; 1 - \frac{m(n)}{2^n}.
$$
Let $\mathsf{coNSIZE}[s(n)]$ and $\mathsf{NSIZE}[s(n)]$ refer to co-nondeterministic circuits and nondeterministic circuits of size $s(n)$, respectively.
More formally, $\mathsf{LB}^1(s_1,s_2,m,n_0)$ is an $\LPV$-sentence capturing the following lower bound statement:
\begin{align*}
&\forall n\in\mathsf{Log}\mathsf{Log}\text{ with }n\ge n_0~\exists C\in\mathsf{coNSIZE}[s_1(n)]~\forall D\in\mathsf{NSIZE}[s_2(n)] \\ 
&\exists m=m(n)\mbox{ distinct $n$-bit strings }x^1,\dots,x^m \mbox{ s.t. } \mathsf{Error}(C,D,x^i)\mbox{ for all }i\in[m],
\end{align*}
where $\mathsf{Error}(C,D,x)$ means that the circuits $C$ and $D$ disagree on the input $x$. This statement can be seen as an average-case form of the $\mathsf{coNP} \nsubseteq \mathsf{NP}/\mathsf{poly}$ conjecture if we let $s_1(n) = n^{O(1)}$, $s_2(n) = n^{\omega(1)}$, and $m(n) = 2^n/n$. (Note that we consider in this section a LogLog formalization, according to the notation explained in \Cref{sec:form_algs_complex}.)

\begin{theorem}[\citep{DBLP:conf/stoc/PichS21, LO23}] \label{thm:LO23CLB}
    Let $d \geq 1$, $\delta > 0$, and $n_0 \geq 1$ be arbitrary parameters, and let $s_1(n) = n^d$, $s_2(n) = 2^{n^\delta}$, and $m(n) = 2^n/n$. Then $\PV$ does not prove the sentence $\mathsf{LB}^1(s_1,s_2,m,n_0)$. 
\end{theorem}

In the remainder of this section, we provide some intuition about the proof of this result.\\

\noindent \textbf{Overview of the Argument.} Suppose, towards a contradiction, that  $\PV \vdash \mathsf{LB}^1(s_1,s_2,m,n_0)$ with parameters as above. The central idea of the argument is that establishing a strong complexity \emph{lower bound} within bounded arithmetic leads to a corresponding complexity \emph{upper bound}. These lower and upper bounds \emph{contradict each other}. Consequently, this contradiction implies the unprovability of the lower bound statement. In a bit more detail, the argument proceeds as follows:
\begin{itemize}
    \item[(\emph{i})] The provability of the average-case lower bound sentence $\mathsf{LB}^1(s_1,s_2,m,n_0)$ implies the provability of a \emph{worst-case} lower bound for $\mathsf{coNSIZE}[n^d]$ against $\mathsf{NSIZE}[2^{n^\delta}]$. We formalize the latter by a sentence $\mathsf{LB}^1_{\mathsf{wst}}(s_1,s_2,n_0)$.
    \item[(\emph{ii})] Given any proof of   $\mathsf{LB}^1_{\mathsf{wst}}(s_1,s_2,n_0)$ in $\PV$, we extract a \emph{complexity upper bound} for an \emph{arbitrary} co-nondeterministic circuit $E_m(x)$ over an input $x$ of length $m$ and of size at most $\poly(m)$. More precisely, we show that there is a deterministic circuit $B_m$ of size $\leq 2^{m^{o(1)}}$ such that $\Pr_{x \sim \{0,1\}^m}[E_m(x) = B_m(x)] \geq 1/2 + 2^{-m^{o(1)}}.$
    \item[(\emph{iii})] We invoke an existing hardness amplification result to conclude that, on any large enough input length $n$, \emph{every} co-nondeterministic circuit $C_n$ of size $\leq n^d$ agrees with some  nondeterministic circuit $D_n$ of size $\leq 2^{n^{\delta}}$ on more than a $1 - 1/n$ fraction of the inputs.
\end{itemize}
Since $\PV$ is a \emph{sound} theory, i.e., every theorem of $\PV$ is a true sentence, Item (\emph{iii}) is in contradiction with the complexity lower bound stated in $\mathsf{LB}^1(s_1,s_2,m,n_0)$. Consequently, $\PV$ does not prove this sentence.\\

The most interesting step of the argument is the proof of Item (\emph{ii}). The key point is that the proof of a lower bound in $\PV$ must be somewhat constructive, in the sense that it not only shows that every small circuit $D$ fails to solve the problem but also produces a string $w$ witnessing this fact. Below we give a simple example of its usefulness, showing a setting where a constructive lower bound yields an upper bound. Note that the application of a witnessing theorem to a LogLog formalization  provides algorithms running in time $\mathsf{poly}(2^n)$. The example provided next shows that this is still useful.

\begin{lemma}[\citep{CLO24b}] \label{thm:UB_from_refuter} Let $L \in \mathsf{NP}$. Suppose that there is a uniform algorithm $R(1^n,D)$ such that, for every co-nondeterministic circuit $D$ on $n$ input variables and of size at most $n^{\log n}$, $R(1^n,D)$ runs in time $2^{O(n)}$ and outputs a string $w \in \{0,1\}^n$ such that $D(w) \neq L(w)$. Then, for every language $L' \in \mathsf{NP}$ and for every constant $\varepsilon > 0$, we have $L' \in \mathsf{DTIME}[2^{n^{\varepsilon}}]$.
\end{lemma}

\begin{proof} Suppose that $L \in \mathsf{NTIME}[n^d]$ for some $d \in \mathbb{N}$. Let $M'$ be a nondeterministic machine that decides $L'$ and runs in time at most $n^{c'}$, where $c' \in \mathbb{N}$. Let $\varepsilon > 0$ be an arbitrary constant. Let $\gamma = \gamma(d, \varepsilon) > 0$ be a small enough constant to be defined later. Finally, let $R$ be the algorithm provided by the hypothesis of the lemma. We show that the following deterministic algorithm $B^\gamma(x)$ decides $L'$ in time  $O(2^{n^{\varepsilon}})$:\\

   \begin{algorithm}[H]
     \SetKwInOut{Input}{Input}
     \caption{Algorithm $B^{\gamma}(x)$ for deciding language $L'$.}\label{algo:ImprovUB}
     \Input{$x \in \{0,1\}^n$ for some $n \geq 1$.}
     Compute the description of a co-nondeterministic circuit $E'$  of size at most $n^{2c'}$ that decides the complement of $L'$\;
     \tcp{In other words, $E'(u) = 1 - L'(u)$ for every string $u \in \{0,1\}^n$.}
       Produce the description  of a co-nondeterministic circuit $D_x(y)$, where $y \in \{0,1\}^{n^\gamma}$, such that $D_x(y)$  ignores its input $y$ and computes according to $E'(x)$\;
    \tcp{While the length of $y$  is smaller than the length of $u$, $D_x$ and $E'$ share the same nondeterministic input string, and $E'$ sets $u$ to be the fixed string $x$.}
    Compute $w = R(1^{n^\gamma}, D_x) \in \{0,1\}^{n^\gamma}$\;
    Determine the bit $b = L(w)$ by a brute force computation, then \Return{$b$}\;
   \end{algorithm}

   \vspace{0.2cm}

First, we argue that $B^\gamma$ decides $L'$. Since $D_x$ is a co-nondeterministic circuit over inputs of length $m \eqdef n^{\gamma}$ and has size at most $n^{2c'} = m^{2c'/\gamma} \leq m^{\log m}$ (for a large enough $m$),   $R(1^{n^\gamma}, D_x)$ outputs a string $w \in \{0,1\}^{n^\gamma}$ such that $L(w) = 1 - D_x(w)$. Consequently,
$$
b = L(w) = 1 - D_x(w) = 1 - E'(x) = 1 - (1 - L'(x))  = L'(x)\;,
$$
i.e., the output bit of $B^\gamma(x)$ is correct.

Next, we argue that $B^\gamma$ runs in time at most $O(2^{n^{\varepsilon}})$. Clearly, Steps $1$--$2$ run in $\mathsf{poly}(n)$ time. Moreover,  Step $3$ runs in time $2^{O(n^{\gamma})}$ under the assumption on the running time of $R(1^{n^\gamma}, D_x)$. This is at most $2^{n^\varepsilon}$ if we set $\gamma \leq \varepsilon/2$. Finally, since $L \in \mathsf{NTIME}[n^d]$, the brute force computation in Step $4$ can be performed in deterministic time $2^{O(\ell^{d})}$ over an input of length $\ell$. Since $\ell = n^{\gamma} = |w|$ in our case, if $\gamma \leq \varepsilon/2d$ we get that Step $4$ runs in time at most $2^{n^{\varepsilon}}$. Overall, if we set $\gamma \eqdef \varepsilon/2d$, it follows that $B^\gamma$ runs in time at most $O(2^{n^\varepsilon})$. This completes the proof that $L' \in \mathsf{DTIME}[2^{n^\varepsilon}]$.
\end{proof}
 
The proof of Item (\emph{ii})  is significantly  more sophisticated, since one does not get an algorithm $R$ as above from a $\PV$ proof of the lower bound sentence $\mathsf{LB}^1(s_1,s_2,m,n_0)$. The argument combines a witnessing theorem for sentences with more than four quantifier alternations and an ingenious technique from \citep{DBLP:journals/jml/Krajicek11} that relies on ideas from the theory of computational pseudorandomness.

\begin{problem}
    Strengthen the unprovability result from \Cref{thm:LO23CLB} in the following directions:
    \begin{itemize}
        \item[\emph{(}a\emph{)}] show that it holds in the polynomial size regime, i.e., with $s_1(n) = n^a$ and for some $s_2(n) = n^b$;
        \item[\emph{(}b\emph{)}] establish the unprovability of worst-case lower bounds against nondeterministic circuits;
        \item[\emph{(}c\emph{)}] show the unprovability of average-case lower bounds against deterministic circuits;
        \item[\emph{(}d\emph{)}] establish the same result with respect to a stronger theory.
    \end{itemize}
\end{problem}

\noindent We refer to \citep{LO23, CLO24b, CLO24a} for some related results and partial progress.

\subsubsection{Extended Frege Lower Bounds}\label{sec:eF_lbs}

This section covers a result on the unprovability of super-polynomial size extended Frege ($\EF$) lower bounds in $\PV$ \citep{krajivcek1990propositional} (see also \citep{MR1241248, buss1990model}). We refer to \Cref{sec:prelim_prop_proofs} for the necessary background. We will also need the definitions and results from \Cref{sec:cuts}.

We adapt the presentation from \citep{krajicek_2019}. Consider the theory $\PV$ and its language $\LPV$. We shall use the following $\LPV$ formulas:
\begin{itemize}
    \item $\mathsf{Sat}(x,y)$: a quantifier-free formula formalizing that $y$ is a satisfying assignment of the Boolean formula $x$;
    \item $\mathsf{Taut}(x) \eqdef \forall y \leq x\,\mathsf{Sat}(x,y)$;
    \item $\Prf_P(x,z)$: a quantifier-free formula formalizing that $z$ is a $P$-proof of $x$.
\end{itemize}

The following lemma is central to the unprovability result. 

\begin{lemma}\label{l:characterisation_no_EF_proof}
    Let $M \models \PV$, and assume that $\phi \in M$ is a propositional formula. The following statements are equivalent:
    \begin{enumerate}
        \item[\emph{(\emph{i})}] There is no $\EF$-proof of $\phi$ in $M$:
        $$ M \models \forall z\,\neg \Prf_{\EF}(\phi,z)\,.
        $$
        \item[\emph{(\emph{ii})}] There is an extension $M' \supseteq M$ \emph{(}also a model of $\PV$\emph{)} in which $\phi$ is falsified:
        $$ M' \models \exists y\,\Sat(\neg \phi, y)\,.
        $$
    \end{enumerate}
\end{lemma}

The proof of \Cref{l:characterisation_no_EF_proof} proceeds by compactness and uses that the correctness of the propositional translation from $\PV$ to $\EF$ (\Cref{sec:prelim_prop_proofs}) is also provable in $\PV$.

\begin{lemma}\label{l:completing_with_EF_proofs}
 Let $M$ be a  nonstandard countable model of $\PV$. Then it has a cofinal extension $M' \supseteq_{\mathsf{cf}} M$ \emph{(}also a model of $\PV$\emph{)} such that every tautology in $M'$ has an $\EF$-proof in $M'$. 
\end{lemma}

The proof of \Cref{l:completing_with_EF_proofs} iterates \Cref{l:characterisation_no_EF_proof} while taking cuts to ensure that the limit extension $M' = \bigcup_i M_i$ (where $M_0=M$) is cofinal in $M$. Since each $M_i \models \PV$ and $\PV$ is universal, we also have $M' \models \PV$. 

We will need the following analogue of \Cref{lemma:majorizing_length_IDelta} for $\PV$.

\begin{fact}\label{fact:majorizing_length_PV}
    Let $M_0$ be a nonstandard countable model of $\PV$. Then there is a \emph{(}countable\emph{)} cut $M$ of $M_0$ that is a \emph{(}nonstandard\emph{)} model of $\PV$ and a length $n \in M$, where $n = |a|$ for some nonstandard $a \in M$, such that for every $b \in M$ we have $M \models |b| \leq n^k$ for some standard number $k$.
\end{fact}

The next result is a consequence of the existence of nonstandard countable models, \Cref{fact:majorizing_length_PV}, and \Cref{l:completing_with_EF_proofs}.

\begin{lemma} \label{lem:model_M_star} There is a model $M^*$ of $\PV$ such that the following properties hold:
    \begin{enumerate}
       \item[\emph{(\emph{i})}]  Any tautology in $M^*$ has an $\EF$-proof in $M^*$.
        \item[\emph{(\emph{ii})}] There is a nonstandard element $a \in M^*$ of length $n \eqdef |a|$ such that for any element $b \in M^*$ there is a standard number $k$ such that $M^* \models |b| \leq n^k$.
    \end{enumerate}
\end{lemma}

\begin{theorem}[Unprovability of super-polynomial size $\EF$ lower bounds in $\PV$ \citep{krajivcek1990propositional}] \label{thm:unprov_EF_PV}
Consider the sentence
$$
\Psi_{\EF} \eqdef \forall x\,\exists\phi \geq x\,[\mathsf{Taut}(\phi) \wedge \forall \pi\,(|\pi| \leq |\phi| \# |\phi| \rightarrow \neg \Prf_{\EF}(\phi,\pi))]\,.\footnote{Recall from \Cref{sec:prelim_complexity} that $x \# y \eqdef 2^{|x| \cdot |y|}$. Consequently, if we let $n = |\phi|$, then the bound $|\pi| \leq |\phi| \# |\phi|$ translates to $|\pi| \leq n \# n$, where $n \# n = 2^{|n| \cdot |n|}$ is of order $n^{\log n}$. The proof of \Cref{thm:unprov_EF_PV} works with any reasonable formalization that refers to a 
 super-polynomial size bound.}
$$
The sentence $\Psi_{\EF}$ is not provable in $\PV$.
\end{theorem}
\begin{proof}
    Suppose $\PV \vdash \Psi_{\EF}$. Let $M^*$, $a$, and $n \eqdef |a|$ be as in \Cref{lem:model_M_star}. Since $\Psi_{\EF}$ holds in $M^*$,  there is a tautology $\phi \in M^*$ with $\phi \geq a$ and consequently $|\phi| \geq n$ such that $\phi$ does not have an $\EF$-proof of size $|\phi|\#|\phi|$ in $M^*$. On the other hand, by the two properties of $M^*$ given by \Cref{lem:model_M_star}, the formula $\phi$ has an $\EF$-proof of size at most $n^k$ for some standard number $k$. Finally, since the element $a$ is nonstandard, we have $n^k \leq n\#n \leq |\phi|\#|\phi|$ in $M^\star$. This contradiction implies that $\PV$ does not prove $\Psi_{\EF}$.
\end{proof}

\begin{problem}
    Show that $\PV$ cannot prove fixed-polynomial size lower bounds on the length of $\EF$ proofs.
\end{problem}

\begin{problem}
    Establish the unprovability of the sentence $\Psi_{\EF}$ in theory $\SOT$.
\end{problem}

\subsection{Connection Between  Upper Bounds and  Lower Bounds}\label{sec:connection_lbs_ubs}

In this section, we explain a result from \citep{BKO20} showing that the unprovability of $\P = \NP$ (\Cref{open:PV_proves_PNP}) is related to the unprovability of circuit lower bounds. For a $\PV$ function symbol $h$ and a circuit size parameter $k \in \mathbb{N}$, consider the sentence
\begin{equation}\label{eq:LB}
\LB(h) \;\eqdef\; \neg \upci[h]\; , \nonumber
\end{equation}
where $\upci[h]$ is the sentence defined in  \Cref{sec:LEARN_unprov}. The sentence  $\LB(h)$ states that the language defined by $h$ is hard on input length $n$ for circuits of size $n^k$ whenever $n$ is sufficiently large. 

\begin{theorem}[Unprovability of {$\P = \NP$} in $\PV$ from the unprovability of  lower bounds in $\PV$ \citep{BKO20}] \label{t:thm_appendix}
If there exists $k \in \mathbb{N}$ such that for no function symbol $h$  theory $\PV$ proves the sentence $\LB(h)$, then for no function symbol $f$  theory $\PV$ proves the sentence $\pnp(f)$.
\end{theorem}

Theorem \ref{t:thm_appendix} shows that if $\PV$ does not prove $n^k$-size lower bounds for a language in $\mathsf{P}$, then $\mathsf{P} \neq \mathsf{NP}$ is consistent with $\PV$. Note that the hypothesis of Theorem \ref{t:thm_appendix} is weaker than the assumption that $\PV$ does not prove that $\mathsf{NP} \nsubseteq \mathsf{SIZE}[n^k]$ for some $k$. 

\begin{proof}[Sketch of the proof of Theorem \emph{\ref{t:thm_appendix}}] 
We proceed in the contrapositive. We formalize in $\PV$ the result that if $\mathsf{P} = \mathsf{NP}$, then for any parameter $k$, $\mathsf{P} \nsubseteq \mathsf{i.o.}\mathsf{SIZE}[n^k]$ (see, e.g., \citep[Theorem 3]{DBLP:conf/coco/Lipton94}). This result combines the collapse of $\mathsf{PH}$ to $\mathsf{P}$ with Kannan's argument \citep{Kan} that $\mathsf{PH}$ can define languages that are almost-everywhere hard against circuits of fixed-polynomial size. Typically, proving this claim requires showing the existence of a truth table of size $2^n$ that is hard against circuits of size $n^k$. However, this result might not be provable in $\PV$.

We address this issue as follows. From the provability in $\PV$ that $\mathsf{P} = \mathsf{NP}$, it follows that for each $i \geq 1$ theory $\mathsf{T}^i_2$ collapses to $\PV$ \citep{KrajicekPT91}. Recall that the dual weak pigeonhole principle ($\mathsf{dWPHP}$) for $\LPV$-functions is provable in $\mathsf{T}^2_2$. Define a $\PV$ function symbol $g$ that takes as input a circuit $C$ of size $n^k$ and outputs the lexicographic first $n^{k + 1}$ bits of the truth table computed by $C$. From $\mathsf{dWPHP}(g)$, we now derive in $\PV$ that the prefix of some truth table is not computable by circuits of size $n^k$, if $n$ is sufficiently large. We can implicitly extend this truth table prefix with zeroes and use the resulting truth table to define a $\PV$-formula $\varphi(x)$ with a constant number of bounded quantifiers that defines a language $L$ that is hard against circuits of size $n^k$, where the hardness is provable in $\PV$. 

Given that the provability in $\PV$ that $\mathsf{P} = \mathsf{NP}$ implies the provability in $\PV$ that $\mathsf{PH}$ collapses to $\mathsf{P}$, it follows that $\varphi(x)$ is equivalent in $\PV$ to the language defined by some $\LPV$-function $h$. In other words, $\PV \vdash \LB(h)$, which completes the proof of Theorem \ref{t:thm_appendix}.
\end{proof}

\citep{CLO24a} shows an example of a simple lower bound that is not provable in $\PV$, under a plausible cryptographic assumption. This indicates that \Cref{t:thm_appendix} might offer a viable approach towards a solution to \Cref{open:PV_proves_PNP}.

\section{Additional Recent Developments}\label{sec:recent_developments}

The provability of the dual Weak Pigeonhole Principle ($\mathsf{dWPHP}$) for polynomial-time functions is closely related to the provability of exponential circuit lower bounds for a language in deterministic exponential time \citep{Jerabek07}. \citep{krajivcek2021small} showed that $\mathsf{dWPHP}$ cannot be proved in $\PV$ under the assumption that $\mathsf{P} \subseteq \mathsf{SIZE}[n^k]$ for some constant $k$.
 \citep{ILW23} established the same unprovability result  assuming
sub-exponentially secure indistinguishability obfuscation and  $\mathsf{coNP} \nsubseteq  \mathsf{i.o.}\mathsf{AM}$.  

\citep{DBLP:conf/stoc/AtseriasBM23} established the unprovability of $\mathsf{NEXP} \subseteq \mathsf{SIZE}[\mathsf{poly}]$ in the theory of bounded arithmetic $\mathsf{V}^0_2$ (not covered in this survey). Interestingly, their approach does not employ a witnessing theorem. It proceeds instead by simulating a   comprehension axiom scheme assuming the provability of the upper bound sentence, eventually relying on an existing lower bound on the provability of the pigeonhole principle.

\citep{CLO24a} systematically  investigates the reverse mathematics of complexity lower bounds. They demonstrated that various lower bound statements in communication complexity, error-correcting codes, and for Turing machines are equivalent to well-studied combinatorial principles, such as the weak pigeonhole principle for polynomial-time functions and its variants. Consequently, complexity lower bounds can be regarded as fundamental axioms with significant implications. They use these equivalences to derive conditional results on the unprovability of simple lower bounds in $\APC_1$.

\citep{CKKOT24} investigates the provability of the circuit size hierarchy in bounded arithmetic,   captured by a sentence $\mathsf{CSH}$ stating that for each $n \geq n_0$, there is a circuit of size $n^a$ that does not admit an equivalent circuit of size $n^b$, where $a > b > 1$ and $n_0$ are fixed. They showed that $\mathsf{CSH}$ is provable in $\mathsf{T}^2_2$, while its provability in $\mathsf{T}^1_2$ implies that $\mathsf{P}^{\mathsf{NP}} \nsubseteq \mathsf{SIZE}[n^{1 + \varepsilon}]$ for some $\varepsilon > 0$. Thus a better proof complexity upper bound for the circuit size hierarchy yields new circuit lower bounds.

\citep{CRT24-Fiat-Shamir} establishes the unprovability of $\mathsf{NP} \neq \mathsf{PSPACE}$ in $\APC_1$ (with a LogLog formalization) under a strong average-case hardness assumption.

\citep{Kra24_Monograph} offers a comprehensive reference on proof complexity generators, whose investigation is closely related to  
$\mathsf{dWPHP}$  and its provability in bounded arithmetic. The theory of proof complexity generators  offers tautologies that serve as potential candidates for demonstrating super-polynomial extended Frege lower bounds and consequently the unprovability of $\P = \mathsf{NP}$ in $\PV$.

We have not covered a number of results connected to the meta-mathematics of complexity lower bounds developed in the context of propositional proof complexity (see, e.g.,~\citep{razborov2015pseudorandom, krajicek_2019, DBLP:conf/coco/AustrinR23, Kra24_Monograph} and references therein). It is worth noting that results on the non-automatability of weak proof systems such as \citep{DBLP:journals/jacm/AtseriasM20, DBLP:conf/stoc/RezendeGNPR021} were made possible thanks to the investigation of the meta-mathematics of proof complexity.

Finally, several other recent papers have investigated directions connected to bounded arithmetic and the meta-mathematics of complexity theory, e.g.,~\citep{DBLP:conf/icalp/PichS22, DBLP:conf/coco/Khaniki22, DBLP:journals/eccc/PichS23,   DBLP:conf/icalp/AKPS24, LLR24}. Due to space constraints, we are not able to cover all recent developments in this survey.

\vspace{-0.1cm}

\small 

\paragraph{Acknowledgements.} I would like to thank Noel Arteche, Jinqiao Hu, Jan Krajíček, Moritz Müller,  Mykyta Narusevych, Ján Pich, and Dimitrios Tsintsilidas for their valuable comments and feedback on an earlier version of this survey. This work received support from the Royal Society University Research Fellowship URF$\setminus$R1$\setminus$191059; the UKRI Frontier Research Guarantee EP/Y007999/1; and the Centre for Discrete Mathematics and its Applications (DIMAP) at the University of Warwick.

\small

\bibliographystyle{alpha}	
\bibliography{references}

\end{document}